\documentclass[preprint,aps,nofootinbib]{revtex4-1}
\usepackage{dcolumn}
\usepackage{bm}
\usepackage{epsfig}
\usepackage{graphicx}
\usepackage{amsmath}
\usepackage{amssymb}
\usepackage{mathrsfs}
\usepackage{color}
\usepackage[usenames,dvipsnames]{xcolor}
\usepackage{hyperref}
\usepackage[caption=false]{subfig}
\usepackage{dsfont}
\allowdisplaybreaks
\usepackage{array}
\usepackage{setspace}
\usepackage{multirow}
\usepackage{makecell}
\setcellgapes{4pt}
\usepackage{enumerate}

\usepackage{amsfonts}

\def\bea{\begin{eqnarray}}
\def\eea{\end{eqnarray}}
\def\bee{\begin{equation}}
\def\ee{\end{equation}}
\def\ba{\begin{align*}}
\def\ea{\end{align*}}

\def\slashchar#1{\setbox0=\hbox{$#1$}           
	\dimen0=\wd0                                 
	\setbox1=\hbox{/} \dimen1=\wd1               
	\ifdim\dimen0>\dimen1                        
	\rlap{\hbox to \dimen0{\hfil/\hfil}}      
	#1                                        
	\else                                        
	\rlap{\hbox to \dimen1{\hfil$#1$\hfil}}   
	/                                         
	\fi}

\begin{document}
	\count\footins = 1000
	\vspace*{.5cm}
	\title{ Minimal Entanglement  and Emergent Symmetries in Low-energy QCD}

	\author{
		\vspace{0.5cm}
		\mbox{Qiaofeng Liu$^{\,a,b}$, Ian Low$^{\,b,c}$ and Thomas Mehen$^{\,a}$}
	}
	\affiliation{
		\vspace*{.5cm}
		$^a$\mbox{{Department of Physics, Duke University, Durham, NC 27708, USA}}\\
		$^b$\mbox{{Department of Physics and Astronomy, Northwestern University, Evanston, IL 60208, USA}} \\
		$^c$\mbox{{High Energy Physics Division, Argonne National Laboratory, Argonne, IL 60439, USA}}
		\vspace{0.5cm}}
	
	\begin{abstract}
		\vspace{0.3cm}
		We study  low-energy scattering of spin-$\frac12$ baryons from the perspective of quantum information science, focusing on the correlation between entanglement minimization and the appearance of accidental symmetries. The baryon transforms as an octet under the $SU(3)$ flavor symmetry and its interactions below the pion threshold are described by contact operators in an effective field theory (EFT) of QCD. Despite there being 64 channels in the 2-to-2 scattering, only six independent operators in the EFT are predicted by $SU(3)$. We show that successive entanglement minimization in $SU(3)$-symmetric channels are correlated with increasingly large emergent symmetries in the EFT. In particular, we identify scattering channels whose entanglement suppression are indicative of emergent $SU(6)$, $SO(8)$, $SU(8)$, and $SU(16)$  symmetries. We also observe the appearance of non-relativistic conformal invariance in channels with unnaturally large scattering lengths. Improved precision from lattice  simulations could help determine the degree of entanglement suppression, and consequently the amount of accidental symmetry, in low-energy QCD.  
		
	\end{abstract}

	\maketitle
	\tableofcontents
	
	\section{Introduction}
	
	Symmetry is among the most fundamental concepts underlying all branches of physics. It is the most powerful guiding principle in formulating laws of nature. Traditionally, there are two views on how symmetry enters into a physical system: those emergent from long-range fluctuations at long distances and those taken as fundamental at very high energy or short distances. The first viewpoint is exemplified in the appearance of scale invariance during a second-order phase transition while the second is embodied in the fact that all known fundamental interactions in nature are dictated by symmetry principles. However, for such a pillar of modern physics there have been very few studies on where the symmetry comes from. Can symmetry be derived from even more fundamental principles?
	
	In pondering the origin of symmetry, a promising line of thought stems from applying tools in quantum information, in particular the concept of entanglement, to study systems with accidental, emergent symmetries in the infrared  \cite{Beane:2018oxh,Low:2021ufv}. The system of interest is low-energy scattering of nucleons (protons and neutrons) which exhibit accidental approximate symmetries not transparent in the fundamental QCD Lagrangian. They include Wigner's supermultiplet $SU(4)_{sm}$ \cite{Wigner:1936dx,Mehen:1999qs}, where the two spin states of protons and neutrons combine into a fundamental representation of $SU(4)_{sm}$, as well as the Schr\"{o}dinger invariance which is the non-relativistic conformal group and the largest symmetry group preserving the Schr\"{o}dinger equation \cite{Mehen:1999nd}. In addition, simulations from lattice QCD  suggest that, for spin-$\frac12$ baryons transforming as the octet of $SU(3)$ flavor symmetry,\footnote{The $SU(3)$ flavor symmetry acts on  the $u$-, $d$- and $s$-quark, which together transform as the fundamental representation of $SU(3)$. It is an exact symmetry of the QCD Lagrangian when the quark masses are neglected, which is a good approximation for  $(u,d,s)$ quarks. Spin-$\frac12$ octet baryons are three-quark bound states of $(u,d,s)$.} there could be an emergent $SU(16)_{sm}$ symmetry analogous to Wigner's $SU(4)_{sm}$, where the two spin states of the eight  baryons furnish a 16-dimensional fundamental representation of $SU(16)_{sm}$ \cite{Wagman:2017tmp}.\footnote{Although Wigner's $SU(4)_{sm}$ can be seen as a consequence of large $N_c$ expansion \cite{su6}, no similar explanation exists for $SU(16)_{sm}$.} Reference~\cite{Beane:2018oxh} made the fascinating observation that the regions of parameter space where the accidental symmetries emerge coincide with regions where the spin entanglement in the 2-to-2 scattering of a proton and a neutron is suppressed, while Ref.~\cite{Low:2021ufv} studied the observation in a quantum information-theoretic setting and showed that the $S$-matrix in the spin space, when viewed as a quantum logic gate, corresponds to an Identity gate in the case of $SU(4)_{sm}$ and $SU(16)_{sm}$ and a SWAP gate in the case of Schr\"{o}dinger symmetry.\footnote{An Identity gate preserves the spin of  qubits (nucleons) while a SWAP gate interchanges the spin of the two qubits. Moreover, these are the only two minimal entanglers for a two-qubit system \cite{Low:2021ufv}.}
	
	These initial findings hint at a rich interplay between entanglement and symmetry and suggest a potentially fruitful probe for the emergence of accidental symmetries using quantum information science. In this work we extend the analyses of Refs.~\cite{Beane:2018oxh,Low:2021ufv} on neutrons ($n$) and protons ($p$) to the eight spin-$\frac12$ baryons transforming as an octet under the $SU(3)$ flavor symmetry: $\{n, p, \Sigma^+, \Sigma^0, \Sigma^-, \Xi^-, \Xi^0, \Lambda\}$. As we will see, because of the rich theoretical structure, the octet baryon offers a fertile playground to further explore the correlation between entanglement minimization and emergent symmetries. Moreover, unlike the scattering of a neutron and a proton, the scattering of two baryons in general can change flavors and the outgoing particles do not have to be the same as the incoming particles. For example, $n$ and  $\Lambda$ have a non-zero probability of scattering into $p$ and  $\Sigma^-$. This feature together with the Pauli exclusion principle, which forces the total wave function of the two incoming/outgoing fermions to be totally antisymmetric,  create a subtle interplay between flavor and spin quantum numbers that is not present in the $np$ scattering.

	We will focus on the very-low-energy scattering of spin-$\frac{1}{2}$ baryons, below the energy threshold for pion production. In this case the process is described by an EFT of QCD using only contact interactions and the  leading order Lagrangian contains only six independent operators  \cite{SW}. The number six is predicted by $SU(3)$ group theory because the product of two octets contains six irreducible representations (irrep): $\mathbf{8}\times \mathbf{8}=\mathbf{27}\oplus \mathbf{10}\oplus \mathbf{\overline{10}}\oplus \mathbf{8}_S\oplus\mathbf{8}_A\oplus \mathbf{1}$. Moreover, since the electric charge and the strangeness quantum number are both conserved in strong interactions, we classify the initial states according to the total electric charge $Q$ and the strangeness $S$, which must remain the same throughout the scattering process. Our analysis uncovers an intriguing pattern that, successive entanglement minimization in $SU(3)$-symmetric and $Q/S$-preserving channels is achieved, an increasingly larger symmetry group appears in the low-energy EFT. We will identify the scattering channels whose spin-entanglement need to be minimized in order to obtain $SU(6)$ spin-flavor symmetries, $SO(8)$ and $SU(8)$ flavor symmetries, as well as $SU(16)_{em}$ spin-flavor symmetry. In addition, in the case of unnatural scattering length, entanglement suppression leads to non-relativistic conformal symmetry in some scattering channels, similar to the $np$ scattering.\footnote{There are other intriguing works on entanglement suppression in scattering of diverse objects, ranging from  black holes \cite{Aoude:2020mlg,Chen:2021huj} to pions \cite{Beane:2021zvo}.} Our findings call for improved precision in lattice QCD simulations of baryon-baryon interactions in the case of natural scattering length, in order to determine the amount of emergent symmetry in low-energy QCD.

	This paper is organized as follows. In Sec.~\ref{entangle} we review measures of entanglement and the definition of entanglement power of operators. We will be applying this measure to find the minimally entangling baryon $S$-matrices. In Sec.~\ref{scatlength} we discuss the momentum expansion of the $S$-matrix for the cases of natural and unnatural scattering length. Succinctly, when the scattering length is natural, i.e., of order the range of the forces, the $S$-matrix is expanded in a power series in the momentum. When the scattering length is unnaturally large the momentum expansion is modified so that  powers of the momentum times scattering length 
	are summed to all orders. Effective field theories of baryons for dealing with these two scenarios are described in Sec.~\ref{EFT}.
	In Sec.~\ref{section-smatrix}, the structure of the baryon-baryon $S$-matrix is presented. Then minimally entangling $S$-matrices and their constraints on phase shifts are considered.
	In Sec.~\ref{section-symmetry} the symmetries of the Lagrangian at each entanglement minimum are considered, followed by a comparison with the numerical simulation in lattice QCD in Sec.~\ref{sect:lqcd}. Our conclusions are given in Sec.~\ref{Conclusion}. This paper has two appendices. Appendix \ref{append:EFT} gives a review of the pionless EFT for nucleon scattering 
	and Appendix \ref{appendix:barybary} gives the composition of baryon states for each irrep of $SU(3)$.

	\section{Entanglement and Entanglement Power}\label{entangle}
	In this section, we briefly review and summarize some of the key concepts in quantum information which are needed in our analysis. We will start with entanglement, which is a property associated with quantum states, and proceed to introduce the entanglement power of an operator.
	
	A quantum state of a system is entangled if it cannot be written as a tensor-product state of its sub-systems. In this case a measurement on a sub-system can modify the state of the rest of the system. Specifically let us consider a bipartite system $\mathcal{H}_{12}$, such as the two-particle system in scattering, whose Hilbert space can be written as a tensor-product space: $\mathcal{H}_{12}=\mathcal{H}_1\otimes\mathcal{H}_2$.
	A state vector $\left|\psi\right\rangle\in \mathcal{H}_{12}$ is  \textit{entangled} if it is not separable,  i.e., there do not exist $\left|\psi_1\right\rangle\in\mathcal{H}_1$ and $\left|\psi_2\right\rangle\in \mathcal{H}_2$ such that $\left|\psi\right\rangle=\left|\psi_1\right\rangle \otimes \left|\psi_2\right\rangle$.
	
	An entanglement measure is a way to quantify the degree of entanglement of any given state. There are multiple entanglement measures. For a bipartite system, the commonly employed von Neumann entropy is defined as
	\bee
	\label{eq:vonent}
	E({\rho}) = -\mathrm{Tr}({\rho}_1\ln{\rho}_1) = -\mathrm{Tr}({\rho}_2\ln{\rho}_2) \ ,
	\ee
	where ${\rho}=|\psi\rangle\langle\psi|$ is the density matrix and ${\rho}_{1(2)}={\rm Tr}_{2(1)}( {\rho})$ is the reduced density matrix obtained after tracing over subsystem 2(1). In computations it is often more convenient to calculate the linear entropy
	\bee\label{linearent}
	E({\rho}) = -\mathrm{Tr}({\rho}_1({\rho}_1-1))= 1- \mathrm{Tr}{\rho}_1^2\  ,
	\ee
	which has the advantage of being a polynomial of ${\rho}_1$. 
	The entanglement measures have the property that they reach zero on a product state $\left\vert\psi_A\right\rangle\otimes\left\vert\psi_B\right\rangle$ and attain a maximum on a maximally entangled state
	
	For a system with two spin-$\frac12$ particles, one often defines the ``computational basis" as $\{\left|\uparrow\uparrow\right\rangle, \left|\uparrow\downarrow\right\rangle, \left|\downarrow\uparrow\right\rangle, \left|\downarrow\downarrow\right\rangle\}$, where $|ij\rangle=|i\rangle_1 \otimes |j\rangle_2$ and an up (down) arrow represents the eigenstate of $\sigma_z$ with eigenvalues $+ 1$ ($-1$). Then for a general normalized spin state
	\bee\label{spin}
	\left|\psi\right\rangle=\alpha\left|\uparrow\uparrow\right\rangle+\beta\left|\uparrow\downarrow\right\rangle+\gamma\left|\downarrow\uparrow\right\rangle+\delta\left|\downarrow\downarrow\right\rangle , \qquad |\alpha|^2+|\beta|^2+|\gamma|^2+|\delta|^2=1 \ ,
	\ee
	its entanglement can be calculated using the linear entropy in Eq.~(\ref{linearent}). The reduced density matrix $\rho_{1}$ for the first spin is 
	\bee
	\rho_{1}=\mathrm{Tr}_2 \left|\psi\right\rangle \left\langle\psi\right| =
	\begin{pmatrix}
		|\alpha|^2+|\beta|^2 & \alpha \gamma^*+\beta\delta^* \\
		\alpha^* \gamma+\beta^*\delta & |\gamma|^2+|\delta|^2 \\
	\end{pmatrix} \ ,
	\ee
	and the entanglement of $\left|\psi\right\rangle$ is therefore
	\begin{align}
	E(\left|\psi\right\rangle) = 1- \mathrm{Tr}_1 \rho_{1}^2 =2\lvert \alpha\delta-\beta\gamma \rvert^2\label{entanglementspin} \ .
	\end{align}
	It is easy to show that (i) if $|\psi\rangle=(a|\!\!\uparrow\rangle_1+b|\!\!\downarrow\rangle_1)\otimes(c|\!\!\uparrow\rangle_2+d|\!\!\downarrow\rangle_2)$ is a product state, $E(\left|\psi\right\rangle)=0$, and (ii) the maximal entanglement is $1/2$, which is the case for the maximally entangled  Bell states $ (\left|\uparrow\uparrow\right\rangle \pm \left|\downarrow\downarrow\right\rangle)/ \sqrt 2$ and  $(\left|\uparrow\downarrow\right\rangle \pm \left|\downarrow\uparrow\right\rangle) / \sqrt 2$.
	
	The linear entropy is related to another measure of spin entanglement, the concurrence $\Delta = \lvert \alpha\delta-\beta\gamma \rvert$  introduced in Ref.~\cite{maximal}. Both measures have the range $(0,1/2)$. In fact, it is possible to show that every entanglement measure for the bipartite qubit system can be expressed in terms of the concurrence $\Delta$ \cite{Low:2021ufv}.

	The entanglement measure quantifies the entanglement in a quantum state $|\psi\rangle$, while the entanglement power measures the ability of a quantum-mechanical operator $U$ to generate entanglement  by averaging over all direct product states it acts on \cite{PhysRevA.63.040304,BallardWu2011}:
	\bee\label{epdef}
	E(U) = \overline{E(U \left| \psi_1 \right>\otimes\left| \psi_2 \right>)} \ ,
	\ee
	where the average is over the state space $\mathcal{H}_1$ and $\mathcal{H}_2$. For qubits we simply average over each Bloch sphere or, equivalently, integrate over all rotations.

	There is, however, a subtlety in the definition of the entanglement power. It is clear that an operator that is local in the product space, $U_1\otimes U_2$, should leave the entanglement of the initial state unchanged, where $U_{1/2}$ is a single-qubit operator acting on $\mathcal{H}_{1/2}$. This leads to the notion of  equivalent classes,
	\bee
	U\sim U^\prime \qquad {\rm if} \qquad U = (U_1\otimes U_2)U^\prime (V_1\otimes V_2) \ .
	\ee
	Modulo  the redundancy in each equivalent class, it was shown in Ref.~\cite{Low:2021ufv} that, for a two-qubit system where $U\in SU(4)$, there are only two operators which minimize the entanglement power in Eq.~(\ref{epdef}): the Identity and the SWAP. They have the following matrix representations in the computational basis,
	\bee
	\bm{1}=\begin{pmatrix}
		\  1 \ &\ 0\ &\ 0\ &\ 0\  \\
		0 & 1 & 0 & 0 \\
		0 & 0 & 1 & 0 \\
		0 & 0 & 0 & 1
	\end{pmatrix},
	\qquad
	\mathrm{SWAP}=\begin{pmatrix}
		\ 1 \ &\ 0\ &\ 0\ &\ 0\ \\0&0&1&0\\0&1&0&0\\0&0&0&1
	\end{pmatrix}\ .
	\ee
	$\mathrm{SWAP}$ interchanges the states of two qubits, which makes it clear that it has minimal entanglement power. 
	It is useful to write SWAP in the product space using Pauli matrices,
	\bee
	\label{eq:swap}
	\mathrm{SWAP} =  (1+\bm{\sigma}\cdot \bm{\sigma})/2\  , \qquad \bm{\sigma}\cdot \bm{\sigma}\equiv \sum_{a}\bm{\sigma}^a\otimes \bm{\sigma}^a \ .
	\ee
	
	In this paper, we primarily study the very low-energy scattering of nucleons and other spin-1/2 baryons, which is dominated by the  $s$-wave for non-identical particles. This implies that the $S$-matrix for $np$ scattering can be written as \cite{Low:2021ufv}
	\bee
	\label{eq:smatrixnp}
	{ S} 
	=  \frac12\left(e^{2i\delta_1}+e^{2i\delta_0}\right) {1} +\frac12\left(e^{2i\delta_1}-e^{2i\delta_0}\right) \text{ SWAP} \ ,
	\ee
where $\delta_{0}$ and $\delta_{1}$ are the scattering phases in the spin singlet and triplet channels, respectively. In general these phases could depend on the center-of-mass momentum $p$. One can see that ${ S} \propto {1}$ if $\delta_0=\delta_1$ and ${ S} \propto {\rm SWAP}$ if $|\delta_0-\delta_1|=\pi/2$. They correspond to the $SU(4)_{sm}$ spin-flavor symmetry and the Schr\"{o}dinger symmetry, respectively. Note that $p \cot \delta_i(p) = -1/a_i +r_{0 i}\, p^2/2 + ...$, so $\delta_0=\delta_1$ can be arranged  by imposing a symmetry that makes the scattering parameters in the two channels equal. Imposing $|\delta_0 - \delta_1| = \pi/2$ for all $p$ is difficult since the $\delta_i(p)$ are complicated functions of $p$ so if the constraint is enforced at one value of momentum it will generically not hold at other values of momentum. However, there is an exception when 
$\delta_0=0$ and $\delta_1=\pm \pi/2$ (or vice versa) for all $p$. $\delta_0=0$ corresponds to $a_0=0$; this is a free theory and the $S$-matrix is trivial. $\delta_1=\pm \pi/2$ corresponds to $1/a_1 = 0$, $r_{1 i}=0$; this corresponds to the unitarity limit \cite{Mehen:1999nd}. Both the free theory and the theory at the unitarity limit are invariant under the Schr\"odinger symmetry, so this is how $|\delta_0-\delta_1 | = \pi/2$ can be realized  in a natural way \cite{Low:2021ufv}.

	We will derive the generalization of Eq.~(\ref{eq:smatrixnp}) for octet baryons, which allows us to directly recognize when the $S$-matrix is minimally entangling, thereby sidestepping the procedure of averaging over initial product states in the definition of entanglement power in  Eq.~(\ref{epdef}). This simplifies our analysis greatly.
	
	\section{Natural and Unnatural Scattering Lengths}
	\label{scatlength}
	In this section we introduce the concept of natural and unnatural scattering lengths in the scattering of non-relativistic particles, focusing on the $s$-wave channel. 
	Assuming a single channel, an equal mass $M$, and energy below any inelastic threshold, the $S$-matrix  for spinless non-relativistic particles can be written as
	\bee
	{S} = e^{2i\delta(p)} = 1+ i\frac{M p}{2\pi} \mathcal{A}\ , \label{eq:smatrix-amplitude}
	\ee
	where $p$ is the center-of-mass momentum, $\delta(p)$ is the phase shift and $\mathcal{A}$ is the scattering amplitude. Rewriting $e^{2i\delta}=(1+i\tan\delta)/(1-i\tan\delta)$ allows us to express
	\bee
	\mathcal{A} = \frac{4\pi}{M}\frac{1}{p\cot\delta-ip} \label{amp}\ .
	\ee
	It has been long known that $p\cot\delta$ admits an expansion in $p^2/\Lambda^2$, where $\Lambda$ is determined by the range of the two-body potential such that it dies off exponentially when the distance $r$ satisfies $r\Lambda > 1$ \cite{GOL64}:
	\bee\label{pcotdelta}
	p\cot\delta = -\frac 1 a + \frac 1 2\, r_0\, p^2 + \dots = -\frac 1 a + \frac 1 2 \Lambda^2 \sum_{n=0}^\infty r_n \left(\frac{p^2}{\Lambda^2}\right)^n.
	\ee
	In the above $a$ is the scattering length,  $r_0$ is the effective range, $r_{1}$ is the shape parameter, etc.   These parameters are measured experimentally. In particular,  $a$ characterizes the effective size of the target and is related to the total cross-section $\sigma_{tot} \to 4\pi a^2$ as $p\to 0$. Equation~(\ref{pcotdelta}) is referred to as the effective range expansion (ERE) \cite{PhysRev.76.38}. 
	
	The $S$-matrix of a free theory is $S=1$ and $\tan\delta=0$ ($\delta=0$); the scattering length $a$ and all $r_i$'s vanish. On the other hand, the interaction is ``maximal'' when $S=-1$ and $\tan\delta=\infty$ ($\delta=\pi/2$); in this case $a\to \infty$ with all other parameters in Eq.~(\ref{pcotdelta}) vanishing and the cross-section is $\sigma_{tot}=4\pi/p^2$, which satisfies the unitarity bound. This is the theory with the largest possible cross section consistent with unitarity and hence this is known as the ``unitarity limit.'' Moreover, at both $\delta=0$ and $\delta=\pi/2$ the theory exhibits Schr\"{o}dinger symmetry, which is sometimes referred to as non-relativistic conformal invariance, since there is no length scale \cite{Mehen:1999nd}.

	In a generic finite-range potential the parameters $r_i$'s in Eq.~(\ref{pcotdelta}) are expected to be of ${\cal O}(1/\Lambda)$ for all $i$.  However $a$ can take any value: it is positive for an attractive potential with bound states, goes to infinity when the bound state is at threshold and has zero binding energy, and becomes negative for weakly attractive potential with resonances (virtual bound states). Recall that bound states are poles in the $S$-matrix residing on the positive imaginary axis on the complex plane and have normalizable wave functions. Resonances are poles in the lower half complex plane whose wave functions are non-normalizable.

	When the scattering length is large compared to $1/\Lambda$, it introduces a subtlety in the perturbative expansion of ${\cal A}$ in $p/\Lambda$. Applying the ERE in Eq.~(\ref{pcotdelta}) to ${\cal A}$ in Eq.~(\ref{amp}), there are two scenarios we would like to highlight:
	\begin{enumerate}[(i)]
		
		\item Natural scattering length: this is when $|a|\alt 1/\Lambda$ (and $|r_i|\alt 1/\Lambda$.) In this case the expansion of  ${\cal A}$ in powers of momentum converges up to $p\sim \Lambda$:
		\bea
		\mathcal{A} &=& - \dfrac{4\pi a}{M}\left[1-i a p + \left(\dfrac{1}{2}a r_0 -a^2\right) p^2 + O(p^3/\Lambda^3)\right] \nonumber\\
		&=& \sum_{n=0}^{\infty} {\cal A}_n \ , \qquad \qquad \qquad {\rm where } \qquad {\cal A}_{n} \sim {\cal O}(p^n) \ . \label{eq:naturalexp}
		\eea
		The leading order (LO) contribution is ${\cal A}_0 \sim {\cal O}(p^0)$.
		
		\item Unnatural scattering length: $|a| \gg 1/\Lambda$, in which case the expansion breaks down when $p\sim 1/|a|$, far below $\Lambda$. It is then necessary to keep $ap$ to all orders when expanding in $p/\Lambda$ \cite{Kaplan:1998tg,Kaplan:1998we}, 
		\begin{align}
		\mathcal{A}&=
		-\frac{4\pi}{M}\frac{1}{(1/a + ip)} \left[1+ \frac{r_0/2}{(1/a + ip)} p^2 + \frac{(r_0/2)^2}{(1/a + ip)^2} p^4 + \dots\right]\nonumber \\
		&={\cal A}_{-1} + \sum_{n=0}^{\infty} {\cal A}_n \ , \qquad \qquad \qquad {\rm where } \qquad {\cal A}_{n} \sim {\cal O}(p^n)\ . \label{eq:unnatural-amplitude-expansion}
		\end{align}
		Notice that the LO term in the amplitude now scales with $p^{-1}$ for $p > 1/|a|$.
	\end{enumerate}

	In low-energy nucleon-nucleon scattering, there is a bound state, the deuteron, in the spin-triplet channel ($^3S_1$) with the scattering length $a_1\approx 5.4$ fm, and a near-threshold virtual bound state in the spin-singlet channel ($^1S_0$) with  $a_0\approx -23.7$ fm. These are (much) larger than the Compton wavelength of pions, $1/m_\pi \approx 1.4$ fm, which sets the range of nuclear  interactions. The fact that the two-nucleon system has  scattering lengths (much) larger than $1/m_\pi$ is the reason why it is often stated that nuclear physics is fined-tuned or unnatural \cite{vanKolck:2020plz}.
	
	\section{EFT for Nucleons and Baryons}\label{EFT}
	
	In the seminal papers \cite{Weinberg:1990rz,Weinberg:1991um} Weinberg proposed using the EFT framework to describe nucleon-nucleon interactions, which has the advantage of employing a systematic expansion with well-defined error estimates for the computation of any observables. Moreover, for the purpose of this work it is  worth emphasizing  that the EFT Lagrangian makes the presence of (approximate) symmetries  transparent. Traditionally the EFT technique requires a separation of scales in the physical system of interest, which in the two-nucleon case is complicated by the presence of unnaturally large scattering lengths.

	We start by considering the effective Lagrangian for a non-relativistic fermion $\psi$ that is invariant under Galilean, parity, and time-reversal symmetries  \cite{Kaplan:1998tg,Kaplan:1998we}:
	\begin{align}
	{\cal L}_{eff}  &= \psi^\dagger \left(i\partial_t + \frac{\nabla^2}{2M} \right) \psi + {C_0} (\psi^\dagger \psi)^2 + \frac{C_2}8 \left[ (\psi \psi)^\dagger(\psi \overset{\text{\small$\leftrightarrow$}}{\nabla}\!\!\!\phantom{1}^2 \psi) + {\rm h.c.}\right] + \cdots\ ,
	\label{eq:Leffnucl}
	\end{align}
	where 
	\bee
	\overset{\text{\small$\leftrightarrow$}}{\nabla}\!\!\!\phantom{1}^2\equiv \overset{\text{\small$\leftarrow$}}{\nabla}\!\!\!\phantom{1}^2 - 2 \overset{\text{\small$\leftarrow$}}{\nabla}\cdot \overset{\text{\small$\rightarrow$}}{\nabla}+\overset{\text{\small$\rightarrow$}}{\nabla}\!\!\!\phantom{1}^2\ ,
	\ee
	is dictated by the Galilean invariance. Terms omitted in ${\cal L}_{eff}$ are contact operators containing more derivatives and/or more fields. In particular, $C_{2n}$ are couplings of contact operators with $2n$ spatial gradients.\footnote{Time derivatives beyond the kinetic term are eliminated in favor of spatial derivatives.} We will focus on the $s$-wave channel and assume the mass $M$ is very large so that relativistic corrections can be neglected. Equation~(\ref{eq:Leffnucl}) can be employed to describe two-nucleon interactions below the pion threshold. In this ``Pionless EFT''  pions are integrated out and the EFT is expected to be valid for $p \alt \Lambda\sim m_\pi$, where $p=\sqrt{M E}$ and $E$ is the momentum and total energy in the center-of-mass frame, respectively.
	
	\begin{figure}[t]
		\begin{center}
			\includegraphics[width=10cm]{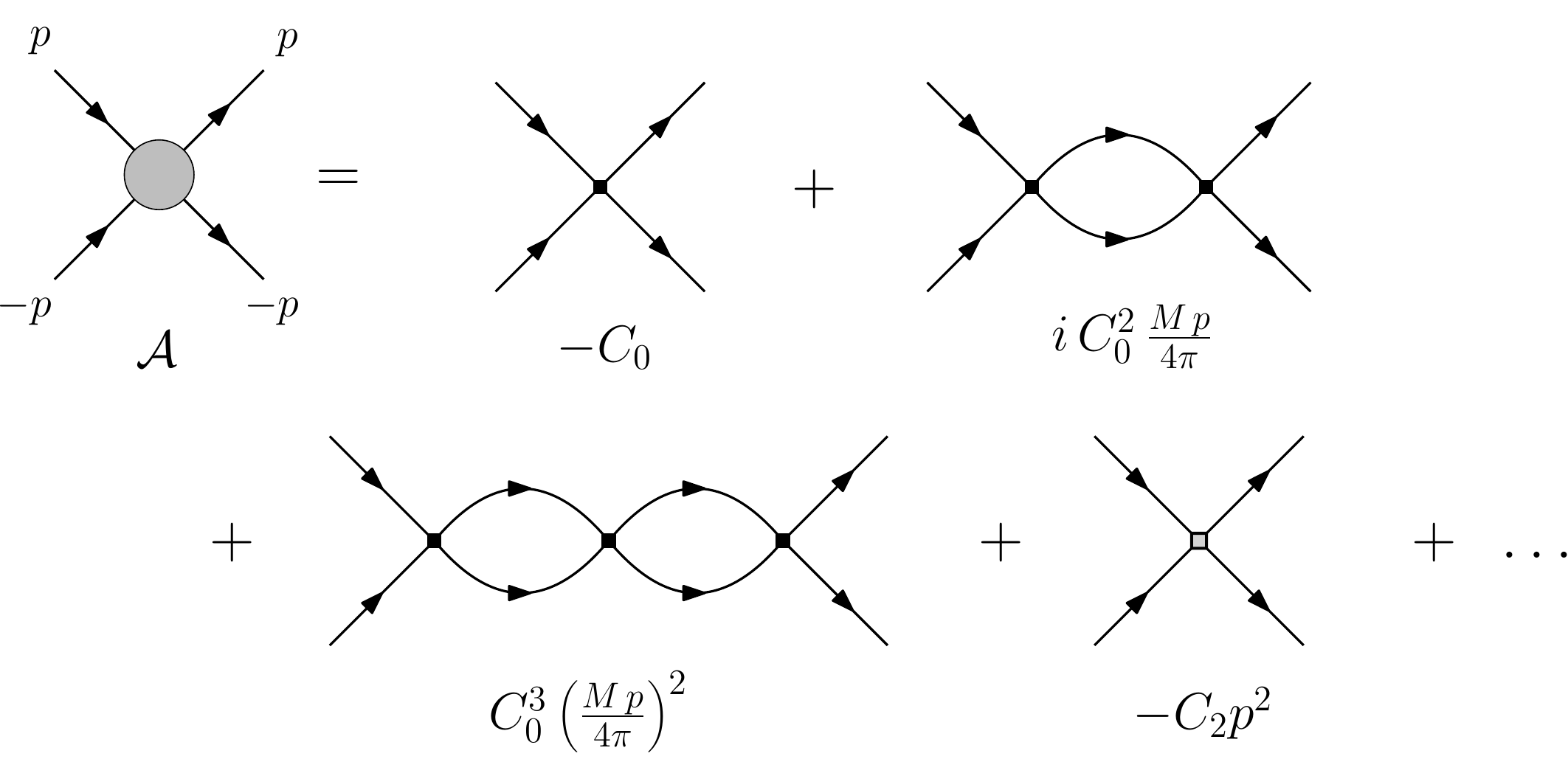}
		\end{center}
		\caption{The first few diagrams contributing to the $s$-wave amplitude in the center-of-mass frame. A solid black vertex represents the $-C_0$ vertex, and a grey vertex represents the $-C_2 \, p^2$ vertex. }
		\label{fig:bubble}
	\end{figure}

	In the EFT the amplitude ${\cal A}$ is usually computed by summing a series of Feynman diagrams to the desired order in the $p/\Lambda$ expansion. This is illustrated in Fig.~\ref{fig:bubble}: at ${\cal O}(p^0)$,  ${\cal A}_0$ comes from a single tree-diagram with one $C_0$ vertex, while  ${\cal A}_1$ arises from a one-loop diagram with two insertions of $C_0$. Going up to ${\cal O}(p^2)$, there are two diagrams, a two-loop diagram with three insertions of $C_0$ and a tree-diagram with one factor of $C_2$, contributing to ${\cal A}_2$. Notice that, in non-relativistic EFT, particles and anti-particles are conserved separately and the only loop-diagrams are the bubble diagrams shown in  Fig.~\ref{fig:bubble}. Importantly, the loop integrals are UV-divergent and  a subtraction scheme is required to remove the UV divergences. In what follows we  summarize the results and relegate the technical details to Appendix \ref{append:EFT}. 
	
	The goal is to have an expansion of the amplitude calculated from EFT that matches ERE. This matching allows us to relate the EFT parameters to physical scattering parameters. Since natural and unnatural scattering have different amplitude expansions, one would also expect different expansion parameters in EFT. This is achieved by carefully selecting the subtraction scheme and the way bubble diagrams are resummed. 
	
	The commonly used minimal subtraction ($MS$) scheme will match the ERE in the natural case. One finds that
	\bee
	C_0 = \frac{4\pi a}M \ , \qquad C_2= C_0\frac{a \, r_0}2 \ .
	\ee
	In general, when the scattering length is of the natural size $a\sim {\cal O}(1/\Lambda)$, $C_{2n} \sim {4\pi}/(M\Lambda^{2n+1})$. But when $a$ is anomalously large, we find instead  $C_{2n} \sim {4\pi a^{n+1}}/(M\Lambda^n)$
	and the perturbative expansion in $p/\Lambda$ breaks down when $1/a \alt p \ll \Lambda$. This is evident in Eq.~(\ref{eq:unnatural-amplitude-expansion}), where the expansion of ${\cal A}$ in ERE starts at ${\cal A}_{-1}\sim 1/p$ for $1/a \alt p$.
	
	It is clear that ${\cal A}_{-1}$, involving a negative power in $p$, can only come from summing over an infinite set of diagrams in perturbation. Moreover, a small coupling must exist to justify the resummation. This is achieved by choosing a different subtraction scheme, the power divergence subtraction (PDS), introduced in Refs.~\cite{Kaplan:1998tg,Kaplan:1998we}, which subtracts both the $1/(D-3)$ and the $1/(D-4)$ poles in the bubble diagram. The PDS subtraction introduces a renormalization scale dependence in the Wilson coefficients $C_{2n}=C_{2n}(\mu)$ and ${\cal A}_{-1}$  arises from summing an infinite number of bubble diagrams with only $C_0$ insertions. In the end,
	\bee
	\label{eq:coc2rg}
	C_0(\mu) = \frac{4\pi}M\left(\frac1{-\mu+1/a}\right) \ ,\qquad C_2(\mu) = \frac{4\pi}M\left(\frac1{-\mu+1/a}\right)^2 \frac{r_0}2 \ .
	\ee
	The $\mu$ dependence makes it possible to establish a new power counting, the KSW-vK counting \cite{Kaplan:1998tg,Kaplan:1998we,vanKolck:1998bw}, which allows for convergence of perturbative expansion over a much larger range of momentum when the scattering length is unnaturally large. This can be seen by taking $\mu \ll 1/|a|$, then  $C_{2n}(\mu)\sim (4\pi/M\Lambda^n)(1/\mu^{n+1})$. Therefore, when $1/|a|  \ll p \alt \Lambda$, we  choose $\mu \sim p$ and
	\bee
	C_{2n}\, p^{2n}\ \sim\ p^{n-1}\ .
	\ee 
	So formally $C_0\sim 1/p$, whose insertions need to be resummed to all orders while insertions of all other $C_{2n>0}$ carry positive powers of $p$ and can be treated perturbatively.

	For neutrons and protons, their masses are almost identical and the EFT has an $SU(2)_I$ isospin symmetry under which they transform as a fundamental representation. Introducing  $p$ and $ n$ as  two-component Pauli spinors representing the two spin states of the proton and neutron, respectively, the LO  interactions in the EFT can be written as \cite{Mehen:1999qs}
	\bee
	\label{eq:Leffnf2}
	{\cal L}_{\rm LO}^{n_f=2} = -\frac12 C_S\, (N^\dagger N)^2  - \frac12 C_T\, (N^\dagger \bm{\sigma} N)\cdot (N^\dagger \bm{\sigma} N) \ , \quad\qquad 
	N=\left( \begin{array}{c} p \\ n \end{array}\right) \ ,
	\ee
	where the Pauli matrices $\bm{\sigma}=(\sigma^1, \sigma^2, \sigma^2)$ only acts on the spin degrees of freedom. $C_S$ and $C_T$ can be expressed in terms of the couplings in the $^1S_0$ and $^3S_1$ channels as $\bar{C}_0=C_S-3C_T$ and $\bar{C}_1=C_S+C_T$, respectively, where  $\bar{C}_i =  (4\pi/M)[1/(-\mu+1/a_i)]$, $i=1,2$, as in Eq.~(\ref{eq:coc2rg}). $M$ is now the nucleon mass and $a_0$ ($a_1$) is the nucleon-nucleon scattering length in the $^1S_0$ ($^3S_1$) channel. Observe that, when $C_T=0$, we have $\bar{C}_0=\bar{C}_1$ which is an indication of Wigner's $SU(4)_{sm}$ spin-isospin symmetry \cite{Wigner:1936dx,Mehen:1999qs}, under which $N$ transforms as a fundamental representation of $SU(4)_{sm}$ and the $C_S$ operator is invariant. On the other hand, $a_i\to \pm \infty$ is a non-trivial fixed point since $\mu(d/d\mu)[\mu \bar{C}_i(\mu)]=0$. In addition to $SU(4)$ symmetry \cite{Mehen:1999qs} in this case one also has the Schr\"{o}dinger symmetry \cite{Mehen:1999nd}.  Ref.~\cite{Beane:2018oxh} pointed out that both cases of emergent symmetries coincide with regions of parameter where the spin-entanglement is suppressed in the 2-to-2 scattering of nucleons. Moreover, the $S$-matrix as a quantum logic gate is the Identity for $SU(4)_{sm}$ and the SWAP  for Schr\"{o}dinger symmetry \cite{Low:2021ufv}.
	
	In low-energy QCD the nucleons are part of the spin-1/2 baryons which transform as an octet under the $SU(3)$ flavor symmetry acting on the light flavor quarks: $(u, d, s)$. The EFT consistent with the $SU(3)_L\times SU(3)_R$ chiral symmetry of QCD is expressed in terms of the $3\times 3$ octet matrix $B$:
	\bee
	\label{eq:baryonmat}
	B=\left(\begin{array}{ccc}
		\Sigma^0/\sqrt{2}+\Lambda/\sqrt{6} & \Sigma^+ & p \\
		\Sigma^- & -\Sigma^0/\sqrt{2}+\Lambda/\sqrt{6} & n \\
		\Xi^- & \Xi^0 & -\sqrt{\frac23} \Lambda 
	\end{array}\right)
	\ee
It is worth pointing out that, in contrast to the similar matrix for the meson octet, the baryon matrix $B$ is not Hermitian because  charge conjugation does not return to the same states; instead it changes baryons to antibaryons. 	
	The LO effective Lagrangian contains six contact operators \cite{SW}:
	\begin{align}
	\label{eq:baryoneft}
	\mathcal{L}_{\rm LO}^{n_f=3}=&-{c_1}\langle B_{i}^{\dagger} B_{i} B_{j}^{\dagger} B_{j}\rangle
	-{c_2}\langle B_{i}^{\dagger} B_{j} B_{j}^{\dagger} B_{i}\rangle
	-{c_3}\langle B_{i}^{\dagger} B_{j}^{\dagger} B_{i} B_{j}\rangle-{c_4}\langle B_{i}^{\dagger} B_{j}^{\dagger} B_{j} B_{i}\rangle \nonumber \\ 
	&-{c_5}\langle B_{i}^{\dagger} B_{i}\rangle\langle B_{j}^{\dagger} B_{j}\rangle-{c_6}\langle B_{i}^{\dagger} B_{j}\rangle\langle B_{j}^{\dagger} B_{i}\rangle \ ,
	\end{align}
	where $\langle\dots\rangle$ denotes the trace over the matrices and $i=1,2$ denotes the two spin states of the baryon. Repeated spin indices are summed over implicitly. It is important to point out that there is a double-trace operator, $\langle B_i^\dagger B_j^\dagger\rangle \langle B_i B_j\rangle$, which in the literature is eliminated using the Cayley-Hamilton identity \cite{SW}:
	\begin{eqnarray}
	\label{eq:cayleyH}
	&&\frac12 \langle B_{i}^{\dagger} B_{i} B_{j}^{\dagger} B_{j}\rangle
	-\frac12 \langle B_{i}^{\dagger} B_{j} B_{j}^{\dagger} B_{i}\rangle
	-\langle B_{i}^{\dagger} B_{j}^{\dagger} B_{i} B_{j}\rangle + \langle B_{i}^{\dagger} B_{j}^{\dagger} B_{j} B_{i}\rangle \nonumber \\
	&=& \frac12 \langle B_{i}^{\dagger} B_{i}\rangle\langle B_{j}^{\dagger} B_{j}\rangle-\frac12 \langle B_{i}^{\dagger} B_{j}\rangle\langle B_{j}^{\dagger} B_{i}\rangle	
	-\frac12 \langle B_i^\dagger B_j^\dagger\rangle \langle B_i B_j\rangle \ .
	\end{eqnarray}
	The fact that there are only six operators at LO is a consequence of $SU(3)$ flavor symmetry group, which will be discussed in the next section.
	
	It is worth mentioning that in the large $N_c$ limit, when the number of color $N_c\to \infty$, the $SU(3)$ flavor symmetry at the quark level is enlarged into the $SU(6)$ quark spin-flavor symmetry \cite{su6}, under which the spin indexes of the $u$, $d$, and $s$ quarks, together with the flavor quantum numbers,  combine to  transform together as a fundamental representation of $SU(6)$. In this case the spin-1/2 and spin-3/2 baryons combine to form a 56-dimensional representation described by a completely symmetric three-index field $\Psi^{\mu \nu \rho}$. The EFT Lagrangian to the lowest order contains only two operators \cite{su6}:
	\bee
	\label{eq:su6eft}
	\mathcal{L}_{int}=
	\left[a\left(\Psi_{\mu \nu \rho}^{\dagger} \Psi^{\mu \nu \rho}\right)^{2}+b \Psi_{\mu \nu \sigma}^{\dagger} \Psi^{\mu \nu \tau} \Psi_{\rho \delta \tau}^{\dagger} \Psi^{\rho \delta \sigma}\right]\ .
	\ee
	In this limit of $SU(6)$ spin-flavor symmetry, the six Wilson coefficients in Eq.~(\ref{eq:baryoneft}) are related to $a$ and $b$ in the above \cite{su6}:
	\begin{align}
	c_1&= -\frac7{27}b \ , & c_2&= \frac19 b \ , \label{eq:su6-1} \\
	c_3&= \frac{10}{81} b \ , & c_4&=-\frac{14}{81}b \ , \label{eq:su6-2}\\
	c_5&=a+\frac29 b \ , & c_6&=-\frac19 b \ . \label{eq:su6-3}
	\end{align}
	We will work out the quantum information-theoretic prediction of the $SU(6)$ symmetry.

	\section{$S$-matrices in Baryon-baryon scattering}\label{section-smatrix}
	
	In this section we study the general $S$-matrix in baryon-baryon scattering, focusing on the interplay between spin and flavor quantum numbers due to the Fermi-Dirac statistics. We establish our notation and formalism by reconsidering the nucleon-nucleon channel and  generalize to octet baryons.  Then we derive conditions for the $S$-matrix to be minimally entangling.
	
	\subsection{Flavors in $np$ scattering}
	
	The $S$-matrix for $np$ scattering in the low-energy limit is dominated by the $s$-wave and contains two channels, $^1S_0$ and $^3S_1$,  and therefore two scattering phases $\delta_0$ and $\delta_1$. In the two-qubit computational basis, $\{|00\rangle, |01\rangle, |10\rangle, |11\rangle\}$, the $S$-matrix can be written as
	\bee
	\label{eq:npchannel}
	{S} = \frac{1-\bm{\sigma}\cdot\bm{\sigma}}{4} e^{2i\delta_0} +\frac{3+\bm{\sigma}\cdot\bm{\sigma}}{4} e^{2i\delta_1}
	\ee
	where $(1-\bm{\sigma}\cdot\bm{\sigma})/4$ and $(3+\bm{\sigma}\cdot\bm{\sigma})/4$ project onto  spin states in the $^1S_0$ and $^3S_1$ subspaces, respectively. The $S$-matrix is manifestly unitary, ${S} {S}^\dagger = 1$. It will be convenient to rewrite Eq.~(\ref{eq:npchannel})  in terms of the Identity and SWAP operators defined in Eq.~(\ref{eq:swap}) \cite{Low:2021ufv},
	\bee
	\label{eq:npsmatnpswap}
	{S} = \frac{1-{\rm SWAP}}{2} \, e^{2i\delta_0} +\frac{1+{\rm SWAP}}{2} \, e^{2i\delta_1} = 1 \left(e^{2i\delta_0}+e^{2i\delta_1}\right) + {\rm SWAP}\left(e^{2i\delta_1}-e^{2i\delta_0}\right)\ ,
	\ee
	from which we see the $S$-matrix is proportional to an Identity gate when $\delta_0=\delta_1$ and the SWAP gate when $|\delta_0-\delta_1|=\pi/2$.
	
	In nature neutrons and protons are almost degenerate in mass, due to the small mass splitting between $u$- and $d$-quarks, which leads to the existence of approximate $SU(2)_I$ isospin symmetry. Under $SU(2)_I$ the proton and neutron are particles with identical isospin quantum number $I=1/2$ but   different $I_3=\pm 1/2$. Using the notation $|I, I_3\rangle$ we have $p=|1/2,1/2\rangle$ and $n=|1/2,-1/2\rangle$. It is instructive to reconsider the $S$-matrix of $NN$ scattering taking into account the isospin invariance, which introduces interesting interplay between flavor and spin quantum numbers due to the Pauli exclusion principle. 
	
	The $S$-matrix for the scattering of two-nucleon $N_{i}(s_i) N_{j}(s_j) \to N_k(s_k) N_l(s_l)$, where $s_i$ represents the spin state of $N_i$,  is now of the tensor-product $({\rm spin})\otimes ({\rm flavor})=({\rm spin}_1\otimes {\rm spin}_2)\otimes ({\rm flavor}_1\otimes {\rm flavor}_2)$:
	\bee
	\label{eq:smatnp}
	{S} = \frac{1-\bm{\sigma}\cdot\bm{\sigma}}{4} \otimes  \frac{3+\bm{\tau}\cdot\bm{\tau}}{4}\, e^{2i\delta_{0}} +\frac{3+\bm{\sigma}\cdot\bm{\sigma}}{4} \otimes  \frac{1-\bm{\tau}\cdot\bm{\tau}}{4}\, e^{2i\delta_{1}}\ ,
	\ee
	where $\bm{\sigma}^a$ and $\bm{\tau}^a$ are Pauli matrices in the spin- and flavor-subspace, respectively. Moreover, because of the Fermi-Dirac statistics, the total isospin $I=1$  projects into the spin singlet $^1S_0$, while the  $I=0$  projects into triplet  $^3S_1$.

	It is worth noting that the $S$-matrix in Eq.~(\ref{eq:smatnp}) is block diagonal in the flavor basis $\{ |pp\rangle, |pn\rangle, |np\rangle, |nn\rangle\}$,
	\bee
	\frac{3+\bm{\tau}\cdot\bm{\tau}}{4} = \begin{pmatrix}
		1  &\ 0 \ &\ 0\  &\ 0 \\
		0 & \ \frac12 &\ \frac12 &\ 0 \\
		0 &\ \frac12 &\ \frac12 &\ 0 \\
		0 &\ 0 &\ 0 &\ 1 \end{pmatrix} \ , \qquad
	\frac{1-\bm{\tau}\cdot\bm{\tau}}{4} = \begin{pmatrix}
		0  &\ 0 \ &\ 0\  &\ 0 \\
		0 & \ \frac12 & -\frac12 &\ 0 \\
		0 & -\frac12 &\ \frac12 &\ 0 \\
		0 &\ 0 &\ 0 &\ 0 \end{pmatrix}          \ .                                                   
	\ee                                                                                                                         
	This is because the electric charge $Q$ is a conserved quantity and commutes with the $S$-matrix. So, alternatively, the $NN$ scattering could be analyzed in individual sectors labeled by the total charge $Q$. For example, focusing on the $Q=+1$ sector which contains $\{|pn\rangle, |np\rangle\}$, the $S$-matrix becomes
	\bee
	\label{eq:smatnpswapnp}
	{S} = \frac{1-\bm{\sigma}\cdot\bm{\sigma}}{4} \otimes  
	\begin{pmatrix} \frac12\ &\ \frac12\ \\
		\frac12\ &\ \frac12\ \end{pmatrix}   \, e^{2i\delta_{0}}                             
	+\frac{3+\bm{\sigma}\cdot\bm{\sigma}}{4} \otimes   \begin{pmatrix} \frac12 & - \frac12 \\
		-\frac12 & \frac12 \end{pmatrix}\, e^{2i\delta_{1}} \ .
	\ee
	In fact, the $S$-matrix in the flavor subspace  is diagonalized in the isospin basis $\{(|pn\rangle+|np\rangle)/\sqrt{2}, (|pn\rangle-|np\rangle)/\sqrt{2}\}$,
	\bee
	\label{eq:smatnpdiag}
	{S} = \frac{1-\bm{\sigma}\cdot\bm{\sigma}}{4} \otimes  
	\begin{pmatrix}  1 \ &\ 0\ \\
		0\ &\ 0 \ \end{pmatrix}   \,    e^{2i\delta_{0}}                         
	+\frac{3+\bm{\sigma}\cdot\bm{\sigma}}{4} \otimes   \begin{pmatrix} 0\ & \ 0\  \\
		0\ & \ 1\  \end{pmatrix}\, e^{2i\delta_{1}} \ ,
	\ee
	where $(|pn\rangle+|np\rangle)/\sqrt{2}$ is in the total isospin $I=1$ and $(|pn\rangle-|np\rangle)/\sqrt{2}$ in the total isospin $I=0$. The projection operators in these two different flavor bases are related by
	\bee
	\begin{pmatrix} \frac12\ &\ \frac12\ \\
		\frac12\ &\ \frac12\ \end{pmatrix} =O_N^\dagger \begin{pmatrix}  1 \ &\ 0\ \\
		0\ &\ 0 \ \end{pmatrix} O_N \ , \qquad 
	\begin{pmatrix} \frac12\ &\ -\frac12\ \\
		-\frac12\ &\ \frac12\ \end{pmatrix} =O_N^\dagger \begin{pmatrix}  0 \ &\ 0\ \\
		0\ &\ 1 \ \end{pmatrix} O_N     \ , 
	\ee
	where
	\bee   
	\label{eq:omatrx}                            
	O_N =      \begin{pmatrix} \frac1{\sqrt{2}}\ &\ \frac1{\sqrt{2}}\ \\
		\frac1{\sqrt{2}}\ &\ -\frac1{\sqrt{2}}\ \end{pmatrix}   \ .                                                  
	\ee                               
	We will see that Eqs.~(\ref{eq:smatnpswapnp}) and (\ref{eq:smatnpdiag}) generalize to baryon-baryon scattering where the isospin $SU(2)_I$ is enlarged to the $SU(3)$ flavor symmetry.

	\subsection{Baryon-baryon scattering}
	\label{sect:subbary}
	
\begin{figure}[t]
		\begin{center}
			\includegraphics[width=6cm]{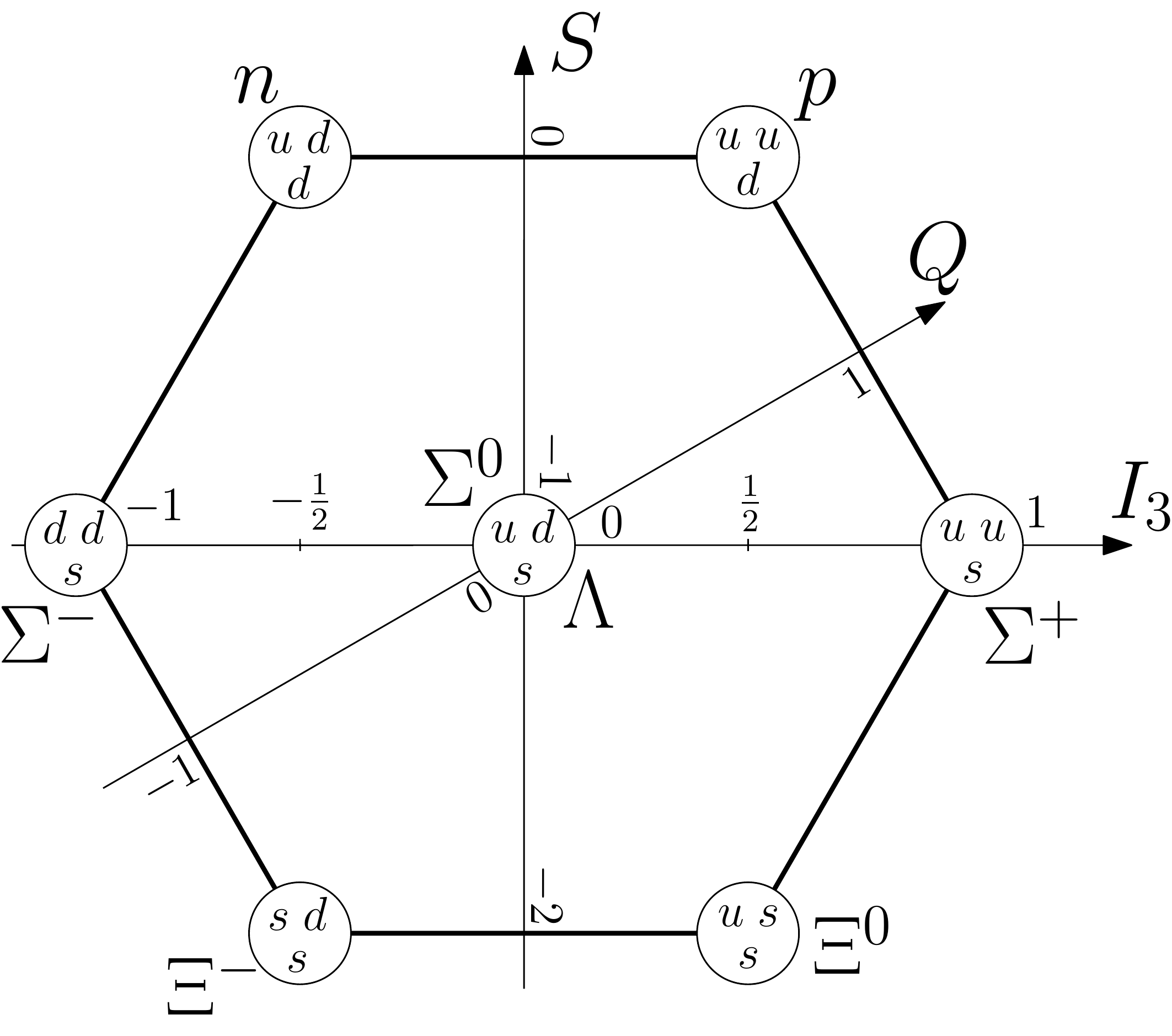}
		\end{center}
		\caption{The spin-1/2 baryons organized in an octet of $SU(3)$ flavor group. The axes, $S$, $Q$, and $I_3$, represent strangeness, charge,  and $I_3$, respectively. }
		\label{fig:octet}
	\end{figure}

	Among the baryons, $np$ scattering is special because it is flavor diagonal, meaning the incoming and outgoing particles are identical and do not change their flavor quantum numbers. This is not the case, in general, for the scattering of other baryons. In Fig.~\ref{fig:octet} we show the lowest-lying spin-1/2 baryon octet, with the states plotted according to their isospin ($I_3$) and strangeness ($S$) quantum numbers.\footnote{In a slight abuse of notation we use $S$ to represent both the strangeness quantum number and the $S$-matrix, with the hope that the distinction is clear from the context.} The electric charge, also shown in Fig.~\ref{fig:octet}, is  $Q=I_3+(S+1)/2$ according to the Gell-Mann-Nishima formula. These baryons furnish an octet representation ($\bm{8}$) under $SU(3)$ flavor group, which then dictates that there are only six scattering phases in the $S$-matrix: $\mathbf{8}\otimes \mathbf{8}=\mathbf{27}\oplus \mathbf{10}\oplus \mathbf{\overline{10}}\oplus \mathbf{8}_S\oplus\mathbf{8}_A\oplus \mathbf{1}$, where $\mathbf{27}$, $\mathbf{8}_S$, $\mathbf{1}$ are $SU(3)$ irreducible representations (irreps) symmetric in the flavor quantum numbers and $\mathbf{10}$, $\mathbf{\overline{10}}$, $\mathbf{8}_A$ are antisymmetric. In addition, the Pauli exclusion principle requires that scattering in the $\mathbf{27}$, $\mathbf{8}_S$, $\mathbf{1}$ must be in the $^1S_0$ channel while $\mathbf{10}$, $\mathbf{\overline{10}}$, $\mathbf{8}_A$ are in  $^3S_1$.  The phases in the $S$-matrix are labeled as $\delta_{\mathbf{27}}$, $\delta_{\mathbf{8_S}} $, $ \delta_{\mathbf{1}} $, $\delta_{\mathbf{10}} $, $ \delta_{\overline{\mathbf{10}}} $, $ \delta_{\mathbf{8_A}}$.  
	
	The presence of  flavor non-diagonal channels  introduces an additional layer of complexity in the $S$-matrix and the interplay between flavor and spin quantum numbers have important consequences when it comes to entanglement minimization. When including the flavor quantum number, two comments are in order: (1) in  a scattering experiment the $|{\rm IN}\rangle$ and $|{\rm OUT}\rangle$ states are usually prepared as a pair of particles with definite flavor quantum numbers in $Q$ and $S$, and (2) in strong interactions $Q$ and $S$ are conserved.
	Due to these considerations, we classify the initial pair of baryons according to the total $Q$ and $S$, as shown in Table.~\ref{sqvalues}. They are divided into various one-dimensional (1-dim), three-dimensional (3-dim) and six-dimensional (6-dim) sectors and an initial pair in a particular $(Q,S)$ sector is only allowed to scatter into other pairs within the sector. The $np$ channel is flavor-diagonal and resides in the $(Q,S)=(1,0)$ sector. The $S$-matrix is block-diagonal in a basis with definite $(Q,S)$ two-baryon states and it is simpler to analyze  each sector individually.

	\begin{table}[t]
	\begin{equation*}
		\begin{array}[t]{>{\centering\arraybackslash$} p{1.5cm} <{$}|>{\centering\arraybackslash$} p{1.0cm} <{$}|>{\centering\arraybackslash$} p{1.0cm} <{$}}
			\hline
			& Q & S\\
			\hline
			n n&0&0\\
			\hline
			n p&1&0\\
			\hline
			p p&2&0\\
			\hline
			n \Sigma^- &-1&-1\\
			\hline
			n \Lambda &&\\
			n\Sigma^0&0&-1\\
			p\Sigma^- &&\\
			\hline
			p \Lambda &&\\
			p \Sigma^0 &1&-1\\
			n\Sigma^+&&\\
			\hline
			p \Sigma^+ &2&-1\\
			\hline
		\end{array}
		\quad
		\begin{array}[t]{>{\centering\arraybackslash$} p{1.5cm} <{$}|>{\centering\arraybackslash$} p{1.0cm} <{$}|>{\centering\arraybackslash$} p{1.0cm} <{$}}
			\hline
			&Q&S\\
			\hline
			\Sigma^-\Sigma^- &-2&-2\\
			\hline
			\Sigma^- \Lambda &&\\
			\Sigma^-\Sigma^0 &-1&-2\\
			n\, \Xi^- &&\\
			\hline
			\Sigma^+\Sigma^- &&\\
			\Sigma^0\Sigma^0 &&\\
			\Lambda \Sigma^0 &&\\
			\Lambda\Lambda &0&-2\\
			n\, \Xi^0 &&\\
			p\, \Xi^- &&\\
			\hline
			\Sigma^+\Lambda &&\\
			\Sigma^+\Sigma^0 &1&-2\\
			p\, \Xi^0 &&\\
			\hline
			\Sigma^+\Sigma^+ &2&-2\\
			\hline
		\end{array}
		\quad
		\begin{array}[t]{>{\centering\arraybackslash$} p{1.5cm} <{$}|>{\centering\arraybackslash$} p{1.0cm} <{$}|>{\centering\arraybackslash$} p{1.0cm} <{$}}
			\hline
			&Q&S\\
			\hline
			\Sigma^- \Xi^- &-2&-3\\
			\hline
			\Sigma^- \Xi^0 &&\\
			\Xi^- \Sigma^0 &-1&-3\\
			\Xi^- \Lambda &&\\
			\hline
			\Xi^- \Sigma^+ &&\\
			\Xi^0 \Lambda &0&-3\\
			\Xi^0 \Sigma^0 &&\\
			\hline
			\Xi^0 \Sigma^+ &1&-3\\
			\hline
			\Xi^-\Xi^-&-2&-4\\
			\hline
			\Xi^- \Xi^0 &-1 &-4\\
			\hline
			\Xi^0 \Xi^0 &0& -4\\
			\hline	
		\end{array}
	\end{equation*}
	\caption{The charge and strangeness of baryon pairs. Strangeness decreases and charge increases from top to bottom.}\label{sqvalues}
\end{table}

\begin{table}[t]
	\begin{tabular}{>{\centering\arraybackslash}m{4cm}|>{\centering\arraybackslash}m{4cm}|>{\centering\arraybackslash}m{5cm}}
		Baryon pairs & symmetric flavor irrep & anti-symmetric flavor irrep\\
		\hline
		\vspace{.07cm} $np,\, \Sigma^-\Xi^-,\, \Sigma^+\Xi^0$ &$\bf{27}$&$\bf{\overline{10}}$\\
		\hline
		$n \Sigma^-,\, p\, \Sigma^+,\, \Xi^-\Xi^0$ &$\bf{27}$&$\bf{10}$\\
	\end{tabular}
	\caption{The baryon pairs in one-dimensional subspaces of $S$ and $Q$ values, and the flavor irreps they sit in. }\label{table:1dsubspaceirreps}
\end{table}
	
	We will start with the 1-dim sectors with non-identical particles,\footnote{For identical particles the particles must be in a spin singlet state and the $S$-matrix simply multiplies this state by a phase.} which includes  the $np$ channel and  involves the $\bm{27}$ (flavor symmetric) and the $\bm{10}$ or $\overline{\bm{10}}$  (flavor antisymmetric) irreps, as shown in Table \ref{table:1dsubspaceirreps}. The $S$-matrix is the same as in Eq.~(\ref{eq:smatnpswapnp}) with $\delta_0=\delta_{\mathbf{27}}$ and $\delta_1=\delta_{\mathbf{10}}$ for the pairs $\{np, \Sigma^-\Xi^-, \Sigma^+\Xi^0\}$, while $\delta_1=\delta_{\overline{\mathbf{10}}}$ for $\{n \Sigma^-, p\, \Sigma^+, \Xi^-\Xi^0\}$.

	For three-dimensional subspaces, the irreps involved are $\mathbf{27}$, $\mathbf{8_S}$, $\mathbf{8_A}$, $\mathbf{10}$ and $\mathbf{\overline{10}}$. (We list all the flavor eigenstates, i.e., states with definite $Q$ and $S,$ within each irrep in Appendix \ref{appendix:barybary}.)  The $S$-matrix can be written in the form, 
	\bee
	{S} = \frac{1-\bm{\sigma}\cdot\bm{\sigma}}{4} \otimes \left( \mathrm{P_\mathbf{27}} \, e^{2i\delta_{\mathbf{27}}} + \mathrm{P_\mathbf{8_S}} \, e^{2i\delta_{\mathbf{8_S}}} \right) +  \frac{3+\bm{\sigma}\cdot\bm{\sigma}}{4} \otimes \left( \mathrm{P_\mathbf{10}} \, e^{2i\delta_{\mathbf{10}}} + \mathrm{P_\mathbf{\overline{10}}} \, e^{2i\delta_{\overline{\mathbf{10}}}} + \mathrm{P_\mathbf{8_A}} \, e^{2i\delta_{\mathbf{8_A}}}   \right), \label{eq:smatrix}
	\ee
	where $\mathrm{P_\mathbf{27}}$ is a projection operator onto the $\bm{27}$ irrep in the flavor space, etc. Again only flavor-symmetric irreps appear in the $^1S_0$ channel; the flavor-antisymmetric irreps reside in the $^3S_1$ channel.
	
	The projection operators can be worked out as follows, using the $(Q,S)=(0,-1)$ sector spanned by $\{\Sigma^0 n, \Sigma^- p, \Lambda n\}$ as an illustration. From Appendix \ref{appendix:barybary}, we see baryon pairs in this sector have components in several irreps:
	\begin{align}
	\label{basis1}
	|\bm{27}\rangle&= \sqrt{\frac23}\ |\Sigma^0 n\rangle_S+\sqrt{\frac13}\ |\Sigma^- p\rangle_S\ , \\ 
	\label{basis11}
	|\bm{27}\rangle^\prime&=-\sqrt{\frac{1}{30}}\ |\Sigma^0 n\rangle_S+\sqrt{\frac{1}{15}}\ |\Sigma^- p\rangle_S + \sqrt{\frac{9}{10}}\ |\Lambda n\rangle_S \ ,  	\\
	|\bf{8_S}\rangle&=  \sqrt{\frac{3}{10}}\ |\Sigma^0 n\rangle_S - \sqrt{\frac35}\ |\Sigma^- p\rangle_S + \sqrt{\frac{1}{10}}\ |\Lambda n\rangle_S \ , \label{basis2} \\
	|\bm{10}\rangle&= \sqrt{\frac23}\ |\Sigma^0 n\rangle_A +\sqrt{\frac13}\ |\Sigma^- p\rangle_A \ , \; \\
	|\mathbf{\overline{10}}\rangle&= -\sqrt{\frac{1}{6}}\ |\Sigma^0 n\rangle_A+\sqrt{\frac{1}{3}}\ |\Sigma^- p\rangle_A + \sqrt{\frac{1}{2}}\ |\Lambda n\rangle_A \ , \; \\
	|\mathbf{8_A}\rangle&= -\sqrt{\frac{1}{6}}\ |\Sigma^0 n\rangle_A + \sqrt{\frac{1}{3}}\ |\Sigma^- p\rangle_A - \sqrt{\frac{1}{2}}\ |\Lambda n \rangle_A \ , \label{basis3} 
	\end{align}
	where the $S$ and $A$ subscripts denote symmetrized and anti-symmetrized flavor quantum numbers states, respectively,
	\bee
	|F_1\ F_2\rangle_{S/A} \equiv \frac1{\sqrt{2}}\left(| F_1\rangle\otimes |F_2\rangle \pm |F_2\rangle \otimes |F_1\rangle\right) \ .
	\ee
	Equations~(\ref{basis1}) - (\ref{basis3}) define an $SU(3)$-flavor basis, in which the flavor part of the $S$-matrix in the  $R$-irrep must be proportional to $e^{2i \delta_R}\ \bm{1}$.\footnote{Recall that the $S$-matrix must be invariant under $SU(3)$ rotations in the respective irrep.} 
	In this basis the projectors in the $S$-matrix are all diagonal, similar to the $np$ scattering in the isospin basis in Eq.~(\ref{eq:smatnpdiag}),
	\begin{eqnarray}
	\label{eq:p26su3}
	\mathrm{P_\mathbf{27}} &=& {\rm Diag}\,(1, 0, 0, 0, 0, 0)\ , \\
	\mathrm{P^\prime_\mathbf{27}} &=& {\rm Diag}\,(0, 1, 0, 0, 0, 0)\ , \\
	\mathrm{P_\mathbf{8_S}} &=&{\rm Diag}\,(0, 0, 1, 0, 0, 0)\ , \\
	\mathrm{P_\mathbf{10}} &=& {\rm Diag}\,(0, 0, 0, 1, 0, 0)\ , \\
	\mathrm{P_\mathbf{\overline{10}}} &=& {\rm Diag}\,(0, 0, 0, 0, 1, 0)\ , \\
	\label{eq:p8asu3}
	\mathrm{P_\mathbf{8_A}} &=& {\rm Diag}\,(0, 0, 0, 0, 0, 1)\ .
	\end{eqnarray}
	These projection operators satisfy the relations: $\sum_{R} \mathrm{P}_R  = 1$  and $\mathrm{P}_R \mathrm{P}_{R'}=\delta_{RR'}\mathrm{P}_R$.

A physical scattering process takes as an initial state one of the states from the flavor eigenbasis consisting of $\{|\Sigma^0 n\rangle$, $|\Sigma^- p\rangle$, $|\Lambda n\rangle$, $|n \Sigma^0\rangle$, $|p \Sigma^-\rangle$, $|n \Lambda\rangle\}$. 
	In this physical basis the projection operator $\mathrm{P_R}$ becomes
	\bee
	\mathrm{P_R}\to O^\dagger\, \mathrm{P_R}\, O\ ,\qquad \qquad
	O= \left(\begin{array}{c|c}\ \frac1{\sqrt{2}} \mathrm{P_S}\ &\  \frac1{\sqrt{2}} \mathrm{P_S}\ \\
		\hline
		\  \frac1{\sqrt{2}} \mathrm{P_A}\ &  \ - \frac1{\sqrt{2}} \mathrm{P_A}  \
	\end{array}\right)\ , \label{eq:physicalbasisprojection}
	\ee                              
	where $O$ is the generalization of Eq.~(\ref{eq:omatrx}) with
	\bee
	\mathrm{P_S} = 
	\begin{pmatrix}
		\sqrt{{2}/{3}} & \sqrt{{1}/{3}} & 0  \\
		-\sqrt{{1}/{30}} & \sqrt{{1}/{15}} & \sqrt{{9}/{10}} \\
		\sqrt{{3}/{10}} & -\sqrt{{3}/{5}} & \sqrt{{1}/{10}} \\ 
	\end{pmatrix} \ , \qquad \quad
	\mathrm{P_A} = 
	\begin{pmatrix}
		\sqrt{{2}/{3}} & \sqrt{{1}/{3}} & 0 \\
		-\sqrt{{1}/{6}} & \sqrt{{1}/{3}} & \sqrt{{1}/{2}} \\
		-\sqrt{{1}/{6}} & \sqrt{{1}/{3}} & -\sqrt{{1}/{2}}
	\end{pmatrix} \ . \label{eq:P_SP_A}
	\ee
	For the 6-dim sector listed in Table \ref{sqvalues}, the construction of the $S$-matrix proceeds similarly. In Appendix \ref{appendix:barybary} we provide the $SU(3)$ basis in all flavor sectors. The basis transformation matrices $\mathrm{P_S}$ and $ \mathrm{P_A}$ in each sector can be read off from the tables in Appendix \ref{appendix:barybary}.

	\subsection{Minimally entangling $S$-matrix}
	
	To study the entanglement property of the $S$-matrix, it will be illuminating to again first consider the nucleon-nucleon channel. In this regard we will rewrite Eq.~(\ref{eq:smatnp})  in terms of the Identity and SWAP operator defined in Eq.~(\ref{eq:swap}),
	\bee
	\label{eq:smatnpswap}
	{S} = \frac{1-{\rm SWAP}}{2} \otimes  \frac{1+\overline{\rm SWAP}}{2}\, e^{2i\delta_{0}} +\frac{1+{\rm SWAP}}{2} \otimes  \frac{1-\overline{\rm SWAP}}{2}\, e^{2i\delta_{1}}\ ,
	\ee
	where $\overline{\rm SWAP}\equiv (1+\bm{\tau}\cdot\bm{\tau})/2$ is the SWAP operator acting in the flavor space. Now if we set $\delta_{0}=\delta_{1}=\delta$,
	\bee
	\label{eq:ssolid}
	S=\frac{e^{2i\delta}}2\,\left(1\otimes 1 -{\rm SWAP}\otimes \overline{\rm SWAP}\right) = e^{2i\delta}\ 1\otimes 1 \ ,
	\ee
	where we have used the fact that ${\rm SWAP} \otimes \overline{\rm SWAP} = - 1\otimes 1$ due to Fermi-Dirac statistics, 
	\bee
	\label{eq:swapsq}
	{\rm SWAP} \otimes \overline{\rm SWAP} \ |N_1, s_1; N_2, s_2\rangle = |N_2, s_2; N_1, s_1\rangle = - |N_1, s_1; N_2, s_2\rangle\ .
	\ee
	So the $S$-matrix is an Identity gate in both the spin and the flavor space!
	Moreover, the emergent $SU(4)_{sm}$ spin-flavor symmetry  is evident since $S$-matrix is proportional to the Identity. Similarly, if $\delta_{1}=\delta_{0}\pm\pi/2=\delta$,
	\bee
	\label{eq:ssolswap}
	S=\frac{e^{2i\delta}}2\,\left({\rm SWAP}\otimes 1- 1\otimes \overline{\rm SWAP}\right) = e^{2i\delta}\ {\rm SWAP}\otimes 1\ ,
	\ee
	where the Fermi-Dirac Statistics again implies
	\bee
	\label{eq:swapide}
	1\otimes \overline{\rm SWAP} \ |N_1, s_1; N_2, s_2\rangle  = - {\rm SWAP}\otimes 1 \ |N_1, s_1; N_2, s_2\rangle \ .
	\ee
	In this case, the $S$-matrix is a SWAP gate in the spin space and an Identity in the flavor space or, equivalently, an Identity in the spin and a SWAP in the flavor.
	
	The discussion above highlights the interplay of entanglement in the spin subspace and in the flavor subspace, in the presence of Fermi-Dirac statistics: the ability of the $S$-matrix to generate entanglement in the spin space is equivalent to the ability to entangle in the flavor space. This is most clear when considering the operator $1\otimes (1+\overline{\rm SWAP})$, which entangles  flavor but not spin. However, using Eq.~(\ref{eq:swapide}) we see
	\bee
	1\otimes (1+\overline{\rm SWAP}) =(1-{\rm SWAP})\otimes 1 \ ,
	\ee
	which now entangles the spin but not flavor.

	More generally, the $S$-matrix in baryon-baryon scattering can be written as 
	\begin{align}
	{S} =\frac{1}{2}  \sum_{\substack{R\in \{\mathbf{27},\mathbf{8_S},\mathbf{1}\}\\R^\prime \in \{\mathbf{10},\mathbf{\overline{10}},\mathbf{8_A}\}}}  \Big[ {1}\otimes \left(\mathrm{P}_R\, e^{2i \delta_R} + \mathrm{P}_{R^\prime}\, e^{2i \delta_{R^\prime}}  \right) - \mathrm{SWAP} \otimes  \left(\mathrm{P}_R\, e^{2i \delta_R} - \mathrm{P}_{R^\prime}\, e^{2i \delta_{R^\prime}}  \right)\Big] \ . \label{eq:Smatrixgeneral}
	\end{align}
	where some irreps are absent in 3-dim and 1-dim sectors. 
	Comparison with Eqs.~(\ref{eq:ssolid}) and (\ref{eq:ssolswap}) suggests that the $S$-matrix in the above equation is minimally entangling if one of the two following conditions are satisfied,
	\begin{align}
	S&\propto \frac{1}2\left(1 \otimes 1 -{\rm SWAP}\otimes \overline{\rm SWAP}\right) = 1\otimes 1\ , \\
	S&\propto \frac12\left({\rm SWAP}\otimes 1- 1\otimes \overline{\rm SWAP}\right) ={\rm SWAP}\otimes 1\ .
	\end{align}
	Let's consider the two possibilities in turn.
	
	The first possibility  is achieved  when
	\begin{align} 
	\label{eq:idensol1}
	&\sum_R\, \mathrm{P}_R\, e^{2i \delta_R} + \sum_{R^\prime}\, \mathrm{P}_{R^\prime}\, e^{2i \delta_{R^\prime}} = e^{2i\delta_I} 1 \ ,\\
	\label{eq:idensol2}
	&\sum_R\, \mathrm{P}_R\, e^{2i \delta_R} - \sum_{R^\prime}\, \mathrm{P}_{R^\prime}\, e^{2i \delta_{R^\prime}} = e^{2i\delta_I}  \overline{\rm SWAP} \ ,
	\end{align}
	where $\delta_I$ is an arbitrary phase shift. 
	To solve for these two relations it is more convenient to choose the $SU(3)$-symmetric basis for the flavor subspace, for examples Eqs.~(\ref{basis1}) - (\ref{basis3}), in which the projection operator takes particularly simple forms, as shown in Eqs.~(\ref{eq:p26su3}) - (\ref{eq:p8asu3}). In this basis the $\overline{\rm SWAP}$ is also simple because the $SU(3)$-symmetric states are  eigenstates of  $\overline{\rm SWAP}$:
	\begin{align}
	&\overline{\rm SWAP}\, | R\rangle =|R\rangle  &{\rm for}& \qquad R\in \{\mathbf{27},\mathbf{8_S},\mathbf{1}\} \ , \\
	& \overline{\rm SWAP}\, | R^\prime\rangle =-|R^\prime\rangle  &{\rm for}& \qquad R^\prime \in \{\mathbf{10},\mathbf{\overline{10}},\mathbf{8_A}\}\ ,
	\end{align}
	where the eigenvalues are $+1$ for the flavor symmetric irreps and $-1$ for the flavor antisymmetric irreps. More explicitly, in the basis of Eqs.~(\ref{basis1}) - (\ref{basis3}), the relations in Eqs.~(\ref{eq:idensol1}) and (\ref{eq:idensol2}) become 
	\begin{eqnarray}
	&&\sum_R\, \mathrm{P}_R\, e^{2i \delta_R} + \sum_{R^\prime}\, \mathrm{P}_{R^\prime}\, e^{2i \delta_{R^\prime}} = e^{2i\delta_I} \left(\begin{array}{c|c}\ {1}_3\ \ &  \\ \hline
	&  {1}_3 \end{array}\right)\ ,\\
	&&\sum_R\, \mathrm{P}_R\, e^{2i \delta_R} - \sum_{R^\prime}\, \mathrm{P}_{R^\prime}\, e^{2i \delta_{R^\prime}} = e^{2i\delta_I} \left(\begin{array}{c|c}\ {1}_3\ \ &  \\ \hline
	&  -{1}_3 \end{array}\right)\ ,
	\end{eqnarray}    
	where ${1}_3$ is a $3\times 3$ Identity matrix. From the matrix representations of $P_{R/R^\prime}$ in  Eqs.~(\ref{eq:p26su3}) - (\ref{eq:p8asu3}),  it is easy to see the solution for the above equation is  when all the scattering phases are equal: 
	\bee
	\label{eq:condid}
	\delta_{R} =\delta_{R^\prime}\equiv \delta_{I} \qquad \Rightarrow \qquad S= e^{2i\delta_I}\ 1 \otimes 1 \ .
	\ee
	So the solution for an Identity gate in the $np$ scattering generalizes to the baryon scattering.

	Next we consider the second possibility, which is reached through the relations
	\begin{align} 
	\label{eq:swapsol1}
	&\sum_R\, \mathrm{P}_R\, e^{2i \delta_R} + \sum_{R^\prime}\, \mathrm{P}_{R^\prime}\, e^{2i \delta_{R^\prime}} =e^{2i\delta_S} \overline{\rm SWAP} \ ,\\
	\label{eq:swapsol2}
	&\sum_R\, \mathrm{P}_R\, e^{2i \delta_R} - \sum_{R^\prime}\, \mathrm{P}_{R^\prime}\, e^{2i \delta_{R^\prime}}  = e^{2i\delta_S} 1  \ .
	\end{align}
	Again in the $SU(3)$-symmetric basis the solution is when
	\bee
	\label{eq:condswap}
	\delta_{R} \equiv \delta_{S} =\delta_{R^\prime}\pm \frac{\pi}2 \qquad \Rightarrow \qquad S= e^{2i\delta_S}\ {\rm SWAP} \otimes 1 \ .
	\ee 
	Again this is a generalization of the solution leading to the SWAP gate in $np$ scattering.                                                                                                                                                                                                                                       
	
	In Table.~\ref{table:minent} we list the explicit solutions for the $S$-matrix to be minimally entangling for each $(Q,S)$ sector.
	
\begin{table}[t]
	\begin{tabular}{>{\centering\arraybackslash}m{7cm}|>{\centering\arraybackslash}m{9cm}}
		$(Q,S)$ sectors & Minimal entanglement conditions \\
		\hline
		$np$&\\
		$\Sigma^-\Xi^-$&  $\delta_{\mathbf{27}} =\delta_{\overline{\mathbf{10}}}\quad$  or  $\quad\delta_{\mathbf{27}} =\delta_{\overline{\mathbf{10}}}\pm \frac{\pi}{2}$ \\
		$\Sigma^+\Xi^0$ \\
		\hline
		$n \Sigma^-$&\\
		$p\, \Sigma^+$&  $\delta_{\mathbf{27}} =\delta_{\mathbf{10}}\quad$  or  $\quad\delta_{\mathbf{27}} =\delta_{\mathbf{10}}\pm \frac{\pi}{2}$ \\
		$\Xi^-\Xi^0$ & \\
		\hline
		$(p \Lambda, \, p \Sigma^0 ,\,  n\Sigma^+ )$ &\\
		$( n\Lambda ,\,n\Sigma^0 $, $p\Sigma^- )$& \\
		$(\Sigma^- \Lambda,\, \Sigma^-\Sigma^0,\, n\, \Xi^-)$ & $\delta_{\mathbf{27}}= \delta_{\mathbf{8_S}} =\delta_{\mathbf{10}} = \delta_{\overline{\mathbf{10}}} = \delta_{\mathbf{8_A}}$ or \\
		$(\Sigma^+\Lambda, \, \Sigma^+\Sigma^0,\, p\, \Xi^0)$ & $\delta_{\mathbf{27}}= \delta_{\mathbf{8_S}} =\delta_{\mathbf{10}} \pm \frac{\pi}{2} = \delta_{\overline{\mathbf{10}}}\pm \frac{\pi}{2} = \delta_{\mathbf{8_A}} \pm \frac{\pi}{2} $  \\
		$(\Sigma^- \Xi^0, \Xi^- \Sigma^0, \Xi^- \Sigma^0)$ & \\
		$(\Xi^- \Sigma^+, \Xi^0 \Lambda, \Xi^0 \Sigma^0)$ & \\
		\hline
		$( \Sigma^+ \Sigma^- ,\, \Sigma^0 \Sigma^0 ,\, \Lambda \Sigma^0 , \, \Xi^- p , \, \Xi^0 n ,\, \Lambda\Lambda )$ & \begin{tabular}{@{}c@{}}\noalign{\smallskip}$\delta_{\mathbf{27}}= \delta_{\mathbf{8_S}} = \delta_{\mathbf{1}} =\delta_{\mathbf{10}} = \delta_{\overline{\mathbf{10}}} = \delta_{\mathbf{8_A}}$ or \\
		$\,\delta_{\mathbf{27}}= \delta_{\mathbf{8_S}} = \delta_{\mathbf{1}} =\delta_{\mathbf{10}} \pm \frac{\pi}{2} = \delta_{\overline{\mathbf{10}}}\pm \frac{\pi}{2} = \delta_{\mathbf{8_A}} \pm \frac{\pi}{2}$\end{tabular}
	\end{tabular}
	\caption{Conditions  in each flavor sector for the $S$-matrix to be minimally entangling. An Identity gate is achieved when all the phases are equal, while a SWAP gate is when the phases differ by $\pi/2$.}\label{table:minent}
\end{table}

	\section{Entanglement Minimum and Symmetry} \label{section-symmetry}

	In this Section we investigate the possible emergent symmetries in the EFT when the $S$-matrix is minimally entangling. The  Lagrangian for the EFT is given in Eq.~(\ref{eq:baryoneft}). It will be convenient to  project the contact operators into irreps of $SU(3)$ flavor symmetry and the corresponding $SU(3)$-symmetric Wilson coefficients are \cite{Wagman:2017tmp}:
	\begin{align}
	C_{\bf 27}&=c_{1}-c_{2}+c_{5}-c_{6} \ , \\
	C_{\bf 8_S}&=-\frac{2 }{3}c_{1}+\frac{2 }{3}c_{2}-\frac{5 }{6}c_{3}+\frac{5 }{6}c_{4}+c_{5}-c_{6} \ ,\\
	C_{\bf 1}&=-\frac{1}{3}c_{1}+\frac{1}{3}c_{2}-\frac{8 }{3}c_{3}+\frac{8 }{3}c_{4}+c_{5}-c_{6} \ , \\
	C_{\bf \overline{10}}&=c_{1}+c_{2}+c_{5}+c_{6}\  ,\\
	C_{\bf 10}&=-c_{1}-c_{2}+c_{5}+c_{6} \ ,\\ 
	C_{\bf 8_A}&=\frac{3 }{2}c_{3}+\frac{3}{2} c_{4}+c_{5}+c_{6}\ .
	\end{align}
	At the leading order in EFT, the relation between the Wilson coefficient $C_i$ and the scattering phase $\delta_i$ is simple and given by Eqs.~(\ref{pcotdelta}) and (\ref{eq:coc2rg}),
	\bee
	p \cot\delta_i = -\left(\mu+\frac{4\pi}{M C_i}\right) \ ,
	\ee 
	where the natural scattering length is recovered by setting $\mu=0$.
	The conditions for a minimally entangling $S$-matrix in Eqs.~(\ref{eq:condid}) and (\ref{eq:condswap}) translate directly into constraints on the Wilson coefficients,
	\begin{align}
	\label{eq:minencon1}
	\delta_R&=\delta_{R'}&  &\implies& & C_R=C_{R'}& &\text{(natural and unnatural)}\ ,\\
	\label{eq:minencon2}
	\delta_R&= \delta_{R'}\pm \dfrac{\pi}{2}& &\implies& & \left\{ \begin{array}{c} C_R = -\dfrac{4\pi}{M\mu}\ ,\ C_{R'}=0  \\
	C_R = 0\ ,\ C_{R'} = -\dfrac{4\pi}{M\mu}\end{array}\right. &&\text{(unnatural)} \ .
	\end{align}
	There are no values of $C_i$ that allow the $\pi/2$ phase difference in the natural scattering length case. Moreover, the values of $C_{R,R'}$ in the unnatural $\pi/2$ phase difference case set $\delta_R=0, \delta_{R'}=\pi/2$ or vice versa. When the scattering phase $\delta=\pi/2$, the scattering length becomes infinite and the channel exhibits the Schr\"{o}dinger symmetry. In this case the channel reaches the unitarity limit.  When the phase $\delta=0$, the $S$-matrix corresponds to a free theory and is also invariant under the Schr\"{o}dinger group.  
	In Table~\ref{table:minentlagrangian} we list the corresponding constraints on the Wilson coefficients for each $(Q,S)$ sector, as deduced from Table~\ref{table:minent}. 
	\begin{table}[t]
	\begin{tabular}{>{\centering\arraybackslash}m{6.6cm}|>{\centering\arraybackslash}m{9.4cm}}
		Flavor subspaces & Minimal entanglement conditions \\
		\hline
		{\renewcommand{\arraystretch}{1.1}		\begin{tabular}{@{}c@{}}
		$np$  \\ 		$\Sigma^-\Xi^-$\\
		$\Sigma^+\Xi^0$\\\end{tabular}}
		&  $c_2=-c_6 \quad$  or  $\quad c_1+c_5 = -\dfrac{2\pi}{M\mu},\,\, c_2+c_6=\pm \dfrac{2\pi}{M\mu}$ \\
		\hline
		{\renewcommand{\arraystretch}{1.1} \begin{tabular}{@{}c@{}}
			\noalign{\smallskip}
		$n \Sigma^-$ \\
		$p\, \Sigma^+$ \\
		$\Xi^-\Xi^0$\end{tabular}}   &  $c_1=c_6 \quad$  or  $\quad -c_2+c_5 = -\dfrac{2\pi}{M\mu},\,\, c_1-c_6=\pm \dfrac{2\pi}{M\mu}$ \\
		\hline
		{\renewcommand{\arraystretch}{1.1}
		\begin{tabular}{@{}c@{}}
		\noalign{\smallskip}
		$(p \Lambda, \, p \Sigma^0 ,\,  n\Sigma^+ )$ \\
		$( n\Lambda ,\,n\Sigma^0 $, $p\Sigma^- )$ \\
		$(\Sigma^- \Lambda,\, \Sigma^-\Sigma^0,\, n\, \Xi^-)$	\\
		$(\Sigma^+\Lambda, \, \Sigma^+\Sigma^0,\, p\, \Xi^0)$  \\
		$(\Sigma^- \Xi^0, \Xi^- \Sigma^0, \Xi^- \Sigma^0)$	\\
		$(\Xi^- \Sigma^+, \Xi^0 \Lambda, \Xi^0 \Sigma^0)$ 
		\\ 
		\end{tabular}} &
		{\renewcommand{\arraystretch}{2}
		 \begin{tabular}{@{}c@{}}
		 $ c_1=-c_2=-\dfrac{1}{2} c_3 = \dfrac{1}{2}c_4 = c_6 $ \; or \\ $c_1=-c_2=-\dfrac{1}{2} c_3 = \dfrac{1}{2}c_4 =-c_5- \dfrac{2\pi}{M\mu}=c_6 \pm \dfrac{2\pi}{M\mu}$ 
		\end{tabular}} 
		\\
		\hline
		$( \Sigma^+ \Sigma^- ,\, \Sigma^0 \Sigma^0 ,\, \Lambda \Sigma^0 , \, \Xi^- p , \, \Xi^0 n ,\, \Lambda\Lambda )$  & 	{\renewcommand{\arraystretch}{1.5}
		\begin{tabular}{@{}c@{}}
		$c_1=c_2=c_3=c_4=c_6=0\quad$ or \\
		$\,c_1=c_2=c_3=c_4=0, c_5=-\dfrac{2\pi}{M\mu}, c_6=\pm \dfrac{2\pi}{M\mu}$ \end{tabular} }
	\end{tabular}
	\caption{Minimal entanglement conditions on Wilson coefficients in each flavor subspace. }\label{table:minentlagrangian}
\end{table}
	
	The previous discussion concerns minimal entanglement in a particular $(Q, S)$ sector and the constraints in Eqs.~(\ref{eq:minencon1}) and (\ref{eq:minencon2}) apply to only a subset of irreps of $\bm{8}\otimes\bm{8}$ that are involved, except for the 6-dim $(Q, S)=(0,-2)$ sector, where all six irreps  enter. In this case, minimal entanglement in the 6-dim sector forces minimal entanglement in all other $(Q, S)$ sectors for scattering of non-identical baryons. This is the the global entanglement minimum and the constraints are
	\begin{align}
	\delta_R=\delta_{R^\prime}\ , \ \ \forall\ {\small R, R^\prime} &\implies& &c_1=c_2=c_3=c_4=c_6= 0\, , \ c_5 \;\text{unconstrained}\ ,  \label{equalphase}\\
	\delta_R=\delta_{R^\prime}\pm \dfrac{\pi}{2} \ , \ \ \forall\  {\small R, R^\prime} &\implies&  &c_1=c_2=c_3=c_4=0\, ,\ c_5=-\frac{2\pi}{M \mu}\, ,\ c_6=\pm \frac{2\pi}{M \mu}\, , \label{diffphase}\end{align}
	where the first line applies to both natural and unnatural cases, and the second line only applies to the unnatural case. The first condition was first reported in Ref.~\cite{Beane:2018oxh} and leads to the emergent $SU(16)_{sm}$ spin-flavor symmetry. The second condition results in the $S$-matrix being a SWAP operator in either the spin or the flavor space and we expect Schr\"{o}dinger symmetry to emerge, similarly to the case of $np$ scattering that was first reported in Ref.~\cite{QiaofengThesis}. 
	
	Next we will use the EFT Lagrangian to investigate the emergent symmetries associated with the conditions listed in Table~\ref{table:minentlagrangian}, starting from the 1-dim sectors and then moving toward the global minimum at the 6-dim sector.
	
	\subsection{Minimal entanglement in 1-dim sectors -- $SU(6)$}
	
	As shown in Table.~\ref{table:1dsubspaceirreps}, there are two different classes of 1-dim sectors: $\{n \Sigma^-$, $p\, \Sigma^+, \Xi^-\Xi^0\}$ involves $\bm{10}$ and $\{np, \Sigma^-\Xi^-, \Sigma^+\Xi^0\}$ contains $\overline{\bm{10}}$. Recall that $\bm{10}$ and $\overline{\bm{10}}$ are irreps of $SU(3)$ related  by complex conjugation, under which $(Q,S+1)\to (-Q, -(S+1))$.   This can also be seen from the ``eight-fold way" in Fig.~\ref{fig:octet}, where the action of complex conjugate gives $n\leftrightarrow \Xi^0$, $p\leftrightarrow \Xi^-$ and $\Sigma^+ \leftrightarrow \Sigma^-$.

	Let us consider $\{n \Sigma^-,p\, \Sigma^+, \Xi^-\Xi^0\}$ first. Minimizing the entanglement in this class requires $C_{{27}}=C_{\overline{{10}}}$. This turns out to be a prediction of the $SU(6)$ spin-flavor symmetry which combines the two spin states with the three quark flavors $(u, d, s)$ into a fundamental representation ($\bm{6}$) of $SU(6)$, whose corresponding Lagrangian in Eq.~(\ref{eq:su6eft}) contains only two parameters $a$ and $b$. Re-expressing the $SU(3)$-symmetric Wilson coefficients using Eqs.~(\ref{eq:su6-1}) - (\ref{eq:su6-3}), we find
	\begin{align}
	&C_{\bf 27} = a-\frac{1}{27}b\ , \qquad C_{\bf 8_S} = a + \frac{1}{3}b\ , \qquad \; C_{\bf 1}=a-\frac{1}{3}b\ ,\\
	&C_{\bf 10} = a+\frac{7}{27}b\ , \qquad C_{\bf \overline{10}}=a-\frac{1}{27}b\ , \qquad
	C_{\bf 8_A} = a+\frac{1}{27}b\ .
	\end{align}
	Thus $SU(6)$ symmetry predicts 
	\bee
	C_{\bf 27} = C_{\bf \overline{10}}\quad \Rightarrow\quad \delta_{\mathbf{27}} = \delta_{\overline{\mathbf{10}}} \ ,
	\ee
	leading to minimal entanglement  in all three channels in $\{n \Sigma^-,p\, \Sigma^+, \Xi^-\Xi^0\}$. In principle there are three other linear relations among the $SU(3)$-symmetric Wilson coefficients following from the $SU(6)$ spin-flavor symmetry, although they do not seem to lead to entanglement suppression in other channels.
	
	On the other hand, requiring minimal entanglement in $\{np, \Sigma^-\Xi^-, \Sigma^+\Xi^0\}$ gives $C_{\bf 27}=C_{\bf 10}$, which is not a prediction of $SU(6)$. However, since  $\{np, \Sigma^-\Xi^-, \Sigma^+\Xi^0\}$ channels are the complex conjugate of $\{\Xi^-\Xi^0 ,p\, \Sigma^+, n \Sigma^- \}$, one can see that $C_{\bf 27}=C_{\bf 10}$ is a prediction of $\overline{SU(6)}$, where the two spin states together with $(u, d, s)$ quarks now transform as the anti-fundamental representation ($\overline{\bm{6}}$) of $SU(6)$.

	\subsection{Minimal entanglement in 3-dim sectors -- $SO(8)$}
	\label{sect:1dso8}
	As seen from Table~\ref{table:minentlagrangian}, in 3-dim flavor sectors the Identity gate is achieved when $c_1=-c_2=-\dfrac{1}{2} c_3 = \dfrac{1}{2}c_4 = c_6 $. Quite unexpectedly, the  Lagrangian in Eq.~(\ref{eq:baryoneft}) simplifies  upon the use of the Cayley-Hamilton theorem in Eq.~(\ref{eq:cayleyH}):
	\bee
	\mathcal{L}  = -(c_1+c_5) \, \langle B_{i}^{\dagger} B_{i}\rangle \langle B_{j}^{\dagger} B_{j}\rangle + c_1 \langle B_{i}^{\dagger} B_{j}^{\dagger}\rangle \langle B_{i} B_{j}\rangle\ .
	\ee
The above Lagrangian  has an emergent $SO(8)$ symmetry. This is the easiest to see by projecting the baryon matrix $B$ in Eq.~(\ref{eq:baryonmat}) into $SU(3)$ generators $T^a$ which satisfy $[T^a, T^b]=if^{abc}T^c$ and $ \mathrm{Tr} \,(T^a T^b) = \dfrac{1}{2} \delta^{ab}$,
\begin{align}
B^a &\equiv {\rm Tr}(BT^a)	\ , \qquad a=1, \cdots, 8\ ,\\
\vec{B} &= (B^1, \cdots, B^8)\nonumber \\
           &=\frac12\left(\Sigma^+ + \Sigma^- , i\Sigma^+ - i\Sigma^- ,  p + \Xi^- , i p - i \Xi^- , n + \Xi^0 , i n - i \Xi^0 , \sqrt{2} \Sigma^0 , \sqrt{2} \Lambda \right)\ , 
\end{align}
where we have chosen $T^a=\lambda^a/2$, where $\lambda^a, a=1, \cdots, 8$ are the famed Gell-Mann matrices for the $SU(3)$ eight-fold way \cite{PhysRev.125.1067}. Then the Lagrangian becomes
	\bee
	\label{eq:c1c5}
	\mathcal{L}  = -2(c_1+c_5) \left(\vec{B}_i^{\dagger}\cdot \vec{B}_i\right)  \left(\vec{B}_j^{\dagger}\cdot \vec{B}_j\right)  + 2c_1 \left(\vec{B}_i^{\dagger}\cdot \vec{B}_j^{\dagger}\right)\left(\vec{B}_i\cdot \vec{B}_j\right),
	\ee
where we recall the $i, j$ indices represents the two spin states of the baryon. In this notation the first operator in Eq.~(\ref{eq:c1c5}) preserves an $SU(8)$ flavor symmetry,
\bee
\label{eq:su8flavor}
	\vec{B} \to \mathcal{U} \vec{B} \ ,\quad \mathcal{U}^\dagger \mathcal{U} = {1}\ .
	\ee
However, the second operator is invariant only under an $SO(8)$ subgroup of $SU(8)$:\footnote{Group generators of $SU(N)$ are $N\times N$ traceless Hermitian matrices, among which the purely imaginary ones are  antisymmetric and hence generators of the $SO(N)$ subgroup.} 
	\bee
	\vec{B} \to \mathcal{O} \vec{B} \ ,\quad \mathcal{O}^T \mathcal{O} = {1}\ , \quad {\rm where}\ \ \mathcal{O} \in SO(8) \ .
	\ee
Thus the Lagrangian in Eq.~(\ref{eq:c1c5}) is invariant under an  emergent $SO(8)$  symmetry.

\begin{table*}[t]
		\begin{tabular}{>{\centering\arraybackslash}m{6.5cm}|>{\centering\arraybackslash}m{9.5cm}}
	Flavor subspace  & Symmetry of Lagrangian \\
	\hline
	{	\begin{tabular}{@{}c@{}}
		$np$  \\ 		$\Sigma^-\Xi^-$\\
		$\Sigma^+\Xi^0$\\ \end{tabular}} &
	\begin{tabular}{@{}c@{}} $SU(6)$ spin-flavor symmetry or \\
	Schr\"odinger symmetry in  $\bm{27}$ and $\bm{\overline{10}}$  irrep channels \end{tabular} \\
	\hline
	{\begin{tabular}{@{}c@{}}
		\noalign{\smallskip}
		$n \Sigma^-$ \\
		$p\, \Sigma^+$ \\
		$\Xi^-\Xi^0$\end{tabular}}  &
	\begin{tabular}{@{}c@{}} conjugate of $SU(6)$ spin-flavor symmetry or \\
	  Schr\"odinger symmetry in $\bm{27}$ and $\bm{10}$ irrep channels 
	\end{tabular} \\
	\hline
	{\renewcommand{\arraystretch}{1.1}
	\begin{tabular}{@{}c@{}}
		\noalign{\smallskip}
		$(p \Lambda, \, p \Sigma^0 ,\,  n\Sigma^+ )$ \\
		$( n\Lambda ,\,n\Sigma^0 $, $p\Sigma^- )$ \\
		$(\Sigma^- \Lambda,\, \Sigma^-\Sigma^0,\, n\, \Xi^-)$	\\
		$(\Sigma^+\Lambda, \, \Sigma^+\Sigma^0,\, p\, \Xi^0)$  \\
		$(\Sigma^- \Xi^0, \Xi^- \Sigma^0, \Xi^- \Sigma^0)$	\\
		$(\Xi^- \Sigma^+, \Xi^0 \Lambda, \Xi^0 \Sigma^0)$ 
		\\ 
	\end{tabular}} &
	\begin{tabular}{@{}c@{}}  $SO(8)$ flavor symmetry or  \\  Schr\"odinger symmetry in $\bm{27}$, $\bf{8_S}$, $\bf{8_A}$, $\bm{10}$ and $\bm{\overline{10}}$ \\ irrep channels 
	\end{tabular} \\
	\hline
	$( \Sigma^+ \Sigma^- ,\, \Sigma^0 \Sigma^0 ,\, \Lambda \Sigma^0 , \, \Xi^- p , \, \Xi^0 n ,\, \Lambda\Lambda )$  &   \begin{tabular}{@{}c@{}} \noalign{\smallskip} $SU(16)$ symmetry \\
	or $SU(8)$ and Schr\"odinger symmetry 	\end{tabular} 
\end{tabular} 
	\caption{\label{table:symmetry}Symmetries predicted by entanglement minimization in each flavor sector.   }
\end{table*}

	\subsection{Minimal entanglement in 6-dim sector  -- $SU(8)$ and $SU(16)$}
	When the global entanglement is minimized, $c_1$--$c_4$ operators all vanish, and only $c_5$ and $c_6$ operators are present, depending on whether the scattering is natural or unnatural. In the case of a natural scattering length, only $c_5$ is non-zero and the Lagrangian is
	\bee
	\mathcal{L}  = -c_5 \, \langle B_{i}^{\dagger} B_{i}\rangle \langle B_{j}^{\dagger} B_{j}\rangle \ .
	\ee
As reported in Ref.~\cite{Beane:2018oxh} this Lagrangian has an $SU(16)$ global symmetry: 
\bee
	{\cal B}=(n_\uparrow, n_\downarrow, p_\uparrow, p_\downarrow, \dots)\ , \qquad \mathcal{L}  = -c_5 \left({\cal B}^\dagger {\cal B}\right)^2 \ .
	\ee
When both $c_5$ and $c_6$ are non-zero, as in the case of an unnatural	scattering length,
\begin{align}
\label{eq:c5c6}
\mathcal{L}  &= -c_5 \, \langle B_{i}^{\dagger} B_{i}\rangle \langle B_{j}^{\dagger} B_{j}\rangle -c_6 \, \langle B_{i}^{\dagger} B_{j}\rangle \langle B_{i}^{\dagger} B_{j}\rangle\nonumber\\
&=- 2 c_5 \left(\vec{B}_i^{\dagger}\cdot \vec{B}_i\right)  \left(\vec{B}_j^{\dagger}\cdot \vec{B}_j\right)  - 2c_6 \left(\vec{B}_i^{\dagger}\cdot \vec{B}_j\right)\left(\vec{B}^\dagger_i\cdot \vec{B}_j\right)\ .
\end{align}
From the reasoning in Section \ref{sect:1dso8}, one sees that the $SU(8)$ symmetry exhibited in Eq.~(\ref{eq:su8flavor}) leaves the above Lagrangian 
 invariant. The $SU(8)$ flavor symmetry is in addition to the non-relativistic conformal invariance that is present when $c_5$ and $c_6$ flows to the UV  fixed point. The full symmetry predictions following from entanglement minimization are listed in Table.~\ref{table:symmetry}.

\section{Results from Lattice QCD}
\label{sect:lqcd}
	
In QCD baryon-baryon interactions were simulated numerically by the NPLQCD Collaboration and four out of the six scattering lengths in $SU(3)$-symmetric channels were evaluated in Ref.~\cite{Wagman:2017tmp} for both  natural scattering length ($\mu=0$) and unnatural scattering length ($\mu=m_\pi$), where $m_\pi=806$ MeV is the mass of the pion in the simulation. More recently new simulations with a more realistic pion mass, $m_\pi=450$ MeV, have appeared in Ref.~\cite{NPLQCD:2020lxg} for three out of the six channels.
			
We list the values of  $C_{R}$ from Refs.~\cite{Wagman:2017tmp,NPLQCD:2020lxg} in Table.~\ref{table:lattice}. Focusing on simulations with  $m_\pi=806$ MeV in Ref.~\cite{Wagman:2017tmp} for now,    we see that, in the case of unnatural scattering length, $C_{\bm{27}} \approx C_{\bm{10}} \approx C_{\overline{\bm{10}}}\approx C_{\bf{8_A}}$. These results imply the presence of $SU(6)$ spin-flavor symmetry, which can then be used to deduce the other two Wilson coefficients $C_{\bf{8_S}}$ and $C_{\bm{1}}$ using Eqs.~(\ref{eq:su6-1}) - (\ref{eq:su6-3}). In the end the Wilson coefficients in all $SU(3)$-symmetric channels are all approximately in the same numerical range, a strong indicator of $SU(16)$ spin-flavor symmetry. On the other hand, in  simulations using more realistic $m_\pi= 450$ MeV in Ref.~\cite{NPLQCD:2020lxg}, only three channels are provided, among which the $SU(6)$ predictions seem to continue to hold up well.
In the case of  natural scattering length, $SU(6)$ spin-flavor symmetry requires $C_{\bm{27}} \approx  C_{\overline{\bm{10}}}$ or $C_{\bm{27}} \approx C_{\bm{10}}$, without which the other two Wilson coefficients cannot be deduced. As can be seen from Table~\ref{table:lattice}, $C_{\bm{27}} \approx  C_{\overline{\bm{10}}}$ seems to work in Ref.~\cite{NPLQCD:2020lxg}, where a lower $m_\pi$ is employed.
\begin{table*}[t]
	\begin{tabular}{>{\centering\arraybackslash}m{3cm}|>{\centering\arraybackslash}m{3cm}|>{\centering\arraybackslash}m{3cm}|>{\centering\arraybackslash}m{3cm}|>{\centering\arraybackslash}m{3cm}}
	& $C_{\bm{27}}$ & $C_{\bm{10}}$ & $C_{\bm{\overline{10}}}$ & $C_{\bf{8_A}}$ \\
	\hline
	natural \cite{Wagman:2017tmp} & $-16.7(2.8)$ & $-50(50)$ & $-11.1(2.5)$ & $-7.7(1.8)$ \\
	\hline
	unnatural \cite{Wagman:2017tmp} & $1.89(4)$ & $1.75(6)$ &  $2.00(8)$ & $2.17(9)$ \\
	\hline 
	natural \cite{NPLQCD:2020lxg} & $-28^{+3}_{-5}$ & - & $-29^{+3}_{-4}$ & $-19^{+1}_{-1}$ \\
	\hline
	unnatural \cite{NPLQCD:2020lxg} & $10.0^{+0.5}_{-0.5}$ & - &  $11.3^{+0.5}_{-0.5}$ & $12.8^{+0.5}_{-0.5}$ \\
	\end{tabular}
		\caption{\label{table:lattice}Wilson coefficients of each irrep, in lattice units, from data in Refs.~\cite{Wagman:2017tmp,NPLQCD:2020lxg}, where $\mu=m_\pi=806$ MeV and 450 MeV, respectively, in the simulations. For $C_{\bm{27}}$ there are several values obtained from different channels and methods listed in Ref.~\cite{NPLQCD:2020lxg}. We picked one with the smallest error bar as a representative. }
\end{table*}
			
At this point it is clear that no firm conclusion can be drawn from the lattice data. It would be desirable to improve on the precision of lattice simulation, as well as computing the Wilson coefficients in more $SU(3)$-symmetric channels, in order to gain further insights on the possible emergent symmetries in low-energy QCD.

	\section{Conclusion}\label{Conclusion}

	In this paper, building upon the work of Refs.~\cite{Beane:2018oxh,Low:2021ufv}, we showed that successive entanglement minimization in $SU(3)$-symmetric channels lead to increasingly large emergent symmetries in low-energy interactions of spin-1/2 baryons, as demonstrated in Table~\ref{table:symmetry}.
	Our findings strongly hint at a new paradigm where symmetry can be considered as the outgrowth of entanglement minimization. With the benefit of hindsight, it may not seem surprising that there is a correlation between entanglement and symmetry since, from the thermodynamic point of view, both  are  related to the presence of ``order" (or the lack thereof) in the physical system.\footnote{Recall that the von Neumann entropy in Eq.~(\ref{eq:vonent}) is an entanglement measure.} This paper represents a first step in establishing a quantitative relation between entanglement and symmetry.
	
	There are many other questions to be answered before we can fully grasp the implications of results in this work. We have considered emergent global symmetries in low-energy QCD. It would be natural to explore similar connections in other physical systems, for example few-body systems in atomic, molecular, and optical (AMO) physics, as well as other types of symmetries such as the gauge symmetry.\footnote{For an exploratory study in this direction, see Ref.~\cite{Cervera-Lierta:2017tdt}.} In addition, global symmetries considered in this work are realized in the Wigner-Weyl mode, where the spectrum of the dynamical system furnishes linear representations of the symmetry group. It would be interesting to understand the Nambu-Goldstone mode, commonly referred to as spontaneous symmetry breaking, from the perspective of quantum information science.\footnote{See Ref.~\cite{Beane:2021zvo} for an initial study.}
	
	Entanglement may also provide a new context to understand another long-standing puzzle in fundamental physics: whether nature is fine tuned. The large scattering length in $np$ scattering in the $^3S_1$ channel gives rise to a near-threshold bound state -- the deuteron -- whose existence is often attributed to fine tuning: the binding energy of deuteron is only 2.2 MeV,  smaller than the typical nuclear binding of ${\cal O}(10)$ MeV. On the other hand, we now know that the large scattering length is associated with the $S$-matrix realizing the SWAP gate. It remains to be seen whether other fine-tuned systems can be understood in a similar fashion.
	
	Last but not least, insights into the precise relation between entanglement and symmetry may in the long run help us devise efficient quantum algorithms to simulate physical systems with a certain type of symmetry. Given the ubiquitous presence of symmetries in nature, this may have broad applications.

	\acknowledgements
	I.L. is supported by the U.S. Department of Energy, Office of Science, Office of Nuclear Physics under Grant No. DE-SC0023522. Q.L. is supported by a University Fellowship from The Graduate School at Northwestern University. T.M. is supported by the U.S. Department of Energy, Office of Science, Office of Nuclear Physics under grant Contract No. DE-FG02-05ER41367.

	\appendix 
	\section{Pionless EFT for nucleons and baryons}
	\label{append:EFT}
	In this appendix, we introduce the pionless EFT, which is essentially a non-relativistic quantum field theory for fermions. The goal is to  explain how to reproduce ERE in Eq.~(\ref{pcotdelta}) in an EFT and discuss  the  power counting scheme  for unnaturally large scattering length.
	
	In a non-relativistic EFT,  the scaling of  $t$ differs from that of  $\vec{x}$ \cite{Hagen:1972pd}:\footnote{To understand the scaling of $t$, recall the non-relativistic dispersion $E= |\vec{p}|^2/(2M)$.}
	\bee
	\vec{x} \to \lambda \vec{x} \ , \qquad \frac{t}M \to \lambda^2\frac{t}{M}  \ , \qquad \psi \to \lambda^{-3/2} \psi \ , 
	\ee
	which leaves the kinetic terms  in Eq.~(\ref{eq:Leffnucl}) invariant. The scaling makes it transparent that the Wilson coefficient of an operator
	with $2n$ spatial derivatives, $C_{2n}$, is of order
	\bee
	\label{eq:c2npower}
	C_{2n}\ \sim \ {\cal O}\left(\frac1{M\Lambda^{2n+1}}\right) \ .
	\ee 
	In this power counting scheme, the leading order diagram is the tree-level diagram with the $C_0$ vertex, as can be seen from Fig.~\ref{fig:bubble}: 
	\bee
	{\cal A}_0 = -C_0 \ . 
	\ee
	It reproduces Eq.~(\ref{eq:naturalexp}) if
	\bee
	\label{eq:appc0}
	C_0 = \frac{4\pi a}{M}\ .
	\ee
	Going to higher orders in Fig.~\ref{fig:bubble} we need to evaluate UV divergent loop integrals, in $D$ spacetime dimensions, of the form \cite{Kaplan:1998tg,Kaplan:1998we}
	\begin{align}
	\label{eq:loopIn}
	I_n &= -i (\mu /2)^{4-D} \int \frac{d^D q}{(2\pi)^D}\frac{i^2\mathbf{q}^{2n}}{(E/2+q_0-\mathbf{q}^2/2M+ i\epsilon)(E/2-q_0-\mathbf{q}^2/2M+i\epsilon)}\nonumber \\
	&= -\frac{(\mu/2)^{4-D}}{(4\pi)^{(D-1)/2}} \Gamma\left(\frac{3-D}{2}\right) M(ME)^n(-ME-i\epsilon)^{(D-3)/2}\ ,
	\end{align}
	and a subtraction scheme is needed.
	Recall  the ERE in the denominator of Eq.~(\ref{amp}) is a polynomial in   $p$, which the EFT aims to reproduce. A subtraction scheme that achieves this goal is the minimal subtraction ($MS$) which removes the $1/(D-4)$ divergence, if any, and leads to \cite{Kaplan:2005es}
	\bee  
	I^{MS}_n = (ME)^n\left(\frac{M}{4\pi}\right) \sqrt{-ME-i\epsilon} = -i\left(\frac{M}{4\pi}\right) p^{2n+1}\ .
	\ee
	The nice feature that the loop momentum $\mathbf{q}$ in $I^{MS}_n$ gets converted to the external momentum $p$ allows one to use the tree-level, on-shell vertex at the operator insertion in the loop diagram. Evaluating the loop diagrams in Fig.~\ref{fig:bubble} and comparing with Eq.~(\ref{eq:naturalexp}) order by order in $p^n$ we arrive at the following relations in the $MS$ scheme  \cite{Kaplan:1998tg,Kaplan:1998we}: 
	\bee
	{\cal A}_0 = -{C_0} \ , \qquad {\cal A}_1 = i\frac{C_0^2\ M p}{4\pi }\ , \qquad {\cal A}_2 = \frac{C_0^3\ M^2 p^2}{16\pi^2 } - {C_2\ p^2} \ , \label{eq:amplitude-lo-natural}
	\ee
	which give
	\bee
	C_0 = \frac{4\pi a}{M} \ , \qquad C_2= C_0\frac{a \, r_0}2 \ .
	\ee
	Notice that in the $MS$ scheme $C_{2n}$ is independent of the renormalization scale $\mu$. Also in this scheme, if $|a|\alt 1/\Lambda$, the perturbative expansion converges up to $\Lambda$ and the power counting in Eq.~(\ref{eq:c2npower}) is reproduced. The EFT in this case could describe a bound state well below the threshold.

	On the other hand, if $a\gg 1/\Lambda \sim 1/m_\pi$, Eq.~(\ref{eq:appc0}) is incompatible with the naive expectation $C_0\sim 1/(M\Lambda)$  and  the perturbative expansion in $C_0$ would not converge in the regime $|1/a| \alt p \alt \Lambda$. The physics behind the breakdown of perturbative expansion in the case of large scattering lengths is the presence of shallow bound states, which manifest themselves as poles in the amplitudes. 
	In this case, one needs to reproduce a different amplitude expansion from ERE, Eq.~(\ref{eq:unnatural-amplitude-expansion}). The leading order term in eq.~(\ref{eq:unnatural-amplitude-expansion}) scales as $p^{-1}$, and since every diagram has a positive order of $p$, $\mathcal{A}_{-1}$ in eq.~(\ref{eq:unnatural-amplitude-expansion}) has to come from an infinite sum of diagrams.
	As pointed out by Weinberg \cite{Weinberg:1990rz,Weinberg:1991um}, poles can be generated by resumming the bubble diagrams (see Fig.~\ref{fig:bubble}) with  any number of $C_0$ insertions:
	\begin{align}
	{\cal A}^{resum} &= -\frac{C_0}{M} +\left(-\frac{C_0}{M}\right) I_0 \left(-\frac{C_0}{M}\right)+\left(-\frac{C_0}{M}\right) I_0 \left(-\frac{C_0}{M}\right) I_0 \left(-\frac{C_0}{M}\right) +\cdots \nonumber \\
	&= \frac{1}{(C_0/M)^{-1} - I_0} \ . \label{eq:C0-resum}
	\end{align}
	Resumming the bubble diagrams is equivalent to solving the Schr\"odinger equation with the potential 
	given by $C_0$. \cite{Weinberg:1990rz,Weinberg:1991um}.

	\begin{figure}[t]
		\begin{center}
			\includegraphics[width=\columnwidth]{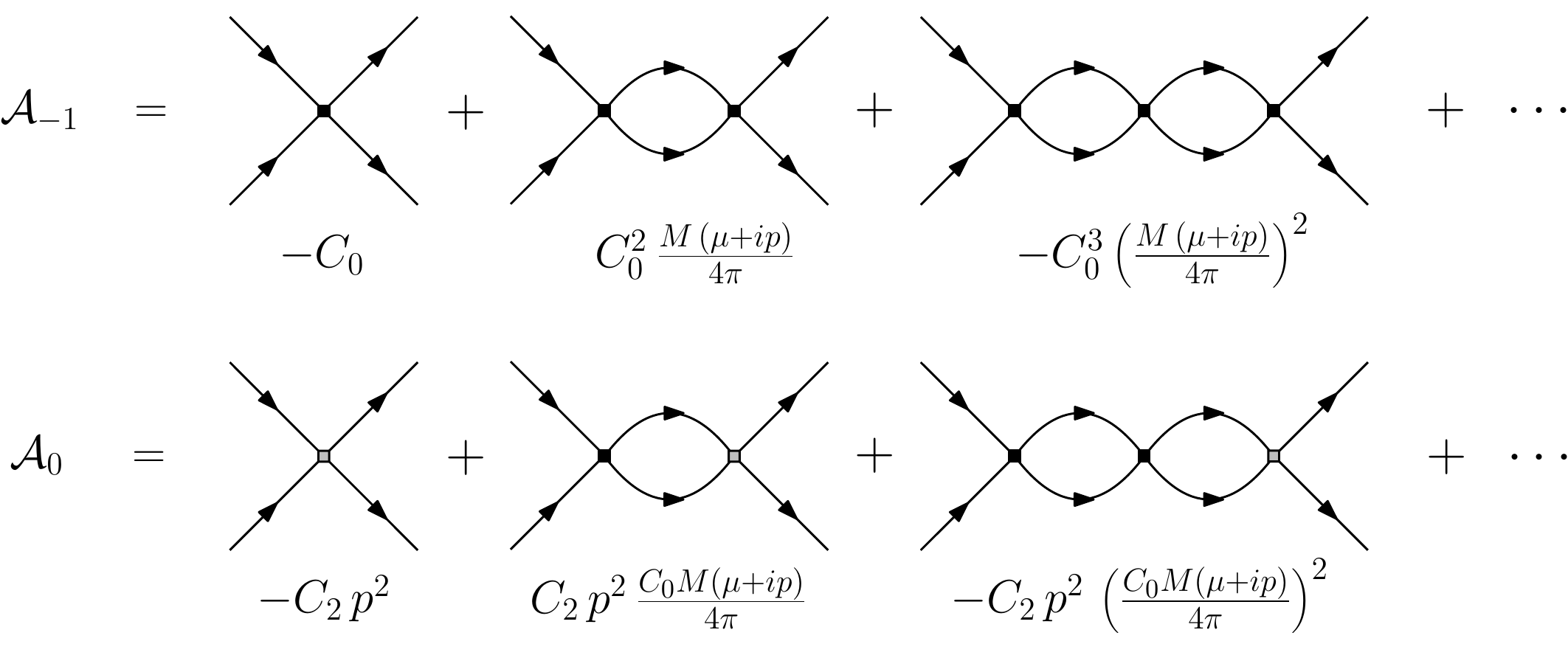}
		\end{center}
		\caption{The diagrams contributing to leading order $s$-wave amplitudes $\mathcal{A}_{-1}$ and $\mathcal{A}_0$ in the center-of-mass frame under PDS scheme. A solid black vertex represents the $-C_0$ vertex, and a grey vertex represents the $-C_2 \, p^2$ vertex. }
		\label{fig:PDSresum}
	\end{figure}
	
    To reproduce eq.~(\ref{eq:unnatural-amplitude-expansion}) in the EFT, one could use a different subtraction scheme, the PDS scheme, and resum all bubble diagrams with insertions of $C_0$, as shown in Fig.~\ref{fig:PDSresum}. This is what is known as the KSW-vK scheme. \cite{vanKolck:1998bw,Bedaque:1997qi,vanKolck:1997ut,Kaplan:1998tg,Kaplan:1998we,Kaplan:2005es}. 
	Under PDS scheme, the $1/(D-3)$ divergence is also removed,
	\bee
	\delta I_n = -\frac{M(ME)^n \mu}{4\pi (D-3)}\ ,
	\qquad
	I_n^{PDS} = -\frac{\mu+ip}{4\pi} M(ME)^n = \frac{-M(\mu+ip)p^{2n} }{4\pi}\ .
	\ee
	The resummed amplitude under PDS scheme is
	\begin{align}
	\mathcal{A}^{resum, PDS} = -\frac{\sum_n C_{2n} p^{2n}}{1+M (\mu+ip)/4\pi \sum_n C_{2n} p^{2n}}\ .
	\label{eq:unnaturalamplitude}
	\end{align}
	To match it to ERE, we consider the quantity 
	\bee
	p \cot\delta = \frac{4\pi}{M\mathcal{A}} +ip = -\frac{4\pi}{M\sum_n C_{2n}P^{2n}} - \mu \ . \label{eq:unnaturalpcoteft}
	\ee
	Comparing the Taylor expansion of eq.~(\ref{eq:unnaturalpcoteft}) and eq.~\ref{pcotdelta}, we get 
	\bee
	C_0 =\frac{4\pi}{M(1/a-\mu)}\ ,\quad  C_2 = \frac{r_0}{2}\frac{C_0}{1/a-\mu} \ . \label{eq:scatteringlengthunnatural}
	\ee
	The leading order terms in amplitude expansion are 
	\bee
	{\cal A}_{-1} = \frac{-C_0}{ 1+ \frac{C_0  M}{4\pi}(\mu+i p)} \ , \quad \qquad {\cal A}_{0} = \frac{-C_2 p^2}{ \left[1+ \frac{C_0 M}{4\pi}(\mu+i p)\right]^2} \ . \label{eq:amplitude-lo-unnatural}
	\ee
It is now evident that	$\mathcal{A}_{-1}$ comprises diagrams with the vertex $C_0$ to all orders, and $\mathcal{A}_0$ comprises diagrams with one $C_2$ insertion and $C_0$ insertions to all orders. 
	The $MS$ scheme results can be recovered by setting $\mu=0$.

	\section{Baryon-baryon scattering channels}
	\label{appendix:barybary}
	The $SU(2)_{spin}\times SU(3)_{flavor}$ symmetry dictates low-energy baryon-baryon interactions. The lowest lying baryons have spin $1/2$ and transform as an eight-dimensional adjoint representation under $SU(3)$ flavor symmetry. Just as two $1/2$ spins can be projected into total spin $S=0$ and $S=1$ states, the 64-dimensional two-baryon flavor space is divided into six irreducible representations (irreps) of $SU(3)$, according to $\bm{8}\otimes\bm{8}=\bm{27}\oplus\bm{10}\oplus\bm{\overline{10}}\oplus\bm{8}_S\oplus\bm{8}_A\oplus\bm{1}$. We list the states of each irrep in the flavor eigenbasis in Tables \ref{channels-27}-\ref{channels-8A}, which are then  used to determine projection matrices  in each flavor sector. It is important to keep in mind that flavors of the two baryons   are symmetrized in $\{\bm{27}, \bm{8}_S,\bm{1}\}$ irreps and anti-symmetrized in $\{\bm{10}, \bm{\overline{10}}, \bm{8}_A\}$ irreps, as we omit symmetric/antisymmetric subscripts in the tables for simplicity. 
	
    We briefly outline the computation process below. 
    Since the Lagrangian is $SU(3)$ invariant, the tensor product decomposition dictates that the Lagrangian is a sum of six $SU(3)$-symmetric operators which mediate 2-to-2 scattering for each of the irreps separately, and whose coefficient will be  denoted as $\{C_{\bm{27}}, C_{\bm{10}}, C_{\bm{\overline{10}}},  C_{\mathbf{8_S}}, C_{\mathbf{8_A}} , C_{\bm 1} \}$.  The $SU(3)$ symmetry further determines that each operator has to be diagonal in the $SU(3)$-symmetric basis.  Since $Q$ and  $S$ are conserved, the Lagrangian can also be split into different $(Q,S)$ sectors, \bee 
    \mathcal{L} = \sum_{Q,S} \mathcal{L}_{Q,S} \ .
    \ee
    Each $\mathcal{L}_{Q,S}$ contains channels from different irreps. One can compute $\mathcal{L}_{Q,S}$ by expanding Eq.~(\ref{eq:baryoneft}) in terms of individual baryon fields and picking out terms in desired $(Q, S)$ sector. $\mathcal{L}_{Q,S}$ is then organized into the diagonal form. For example, the Lagrangian involving the $\{\Sigma^0 n, \Sigma^- p, \Lambda n\}$ sector is diagonalized as 
    \begin{align}
    \mathcal{L}_{Q=0,S=-1} &=- (c_{1}-c_{2}+c_{5}-c_{6}) (\psi_{1}^\dagger \psi_{1} + \psi_{2}^{\dagger} \psi_{2}) \nonumber \\
    &- \left(-\frac{2}{3}c_{1}+\frac{2}{3}c_{2}-\frac{5 }{6}c_{3}+\frac{5}{6}c_{4}+c_{5}-c_{6}\right) \psi_{3}^\dagger \psi_{3}\nonumber \\
    & - \left(c_{1}+c_{2}+c_{5}+c_{6}\right) (\psi_{4}^{(1)\,\dagger} \psi_{4}^{(1)} +\psi_{4}^{(2)\,\dagger} \psi_{4}^{(2)} + \psi_{4}^{(3)\,\dagger} \psi_{4}^{(3)}) \nonumber  \\
    & -\left(-c_{1}-c_{2}+c_{5}+c_{6}\right) (\psi_{5}^{(1)\,\dagger} \psi_{5}^{(1)} +\psi_{5}^{(2)\,\dagger} \psi_{5}^{(2)} + \psi_{5}^{(3)\,\dagger} \psi_{5}^{(3)})\nonumber \\
    & - \left(\frac{3 }{2}c_{3}+\frac{3}{2} c_{4}+c_{5}+c_{6}\right) (\psi_{6}^{(1)\,\dagger} \psi_{6}^{(1)} +\psi_{6}^{(2)\,\dagger} \psi_{6}^{(2)} + \psi_{6}^{(3)\,\dagger} \psi_{6}^{(3)})  \ ,
    \end{align}
    where
    \begin{align}
    \psi_{1} &= \sqrt{\frac23}\left(\frac{\Sigma^0_\uparrow n_\downarrow - \Sigma^0_\downarrow n_\uparrow}{\sqrt{2}}\right)+ \sqrt{\frac13}\left(\frac{\Sigma^-_\uparrow p_\downarrow - \Sigma^-_\downarrow p_\uparrow}{\sqrt{2}}\right)\ , \\
    \psi_{2} & = -\sqrt{\frac{1}{30}}\ \left(\frac{\Sigma^0_\uparrow n_\downarrow - \Sigma^0_\downarrow n_\uparrow}{\sqrt{2}}\right) +\sqrt{\frac{1}{15}}\ \left(\frac{\Sigma^-_\uparrow p_\downarrow - \Sigma^-_\downarrow p_\uparrow}{\sqrt{2}}\right) + \sqrt{\frac{9}{10}}\ \left(\frac{\Lambda_\uparrow n_\downarrow - \Lambda_\downarrow n_\uparrow}{\sqrt{2}}\right) \ , \\
    \psi_{3} &= \sqrt{\frac{3}{10}}\ \left(\frac{\Sigma^0_\uparrow n_\downarrow - \Sigma^0_\downarrow n_\uparrow}{\sqrt{2}}\right) - \sqrt{\frac35}\ \left(\frac{\Sigma^-_\uparrow p_\downarrow - \Sigma^-_\downarrow p_\uparrow}{\sqrt{2}}\right)  + \sqrt{\frac{1}{10}}\ \left(\frac{\Lambda_\uparrow n_\downarrow - \Lambda_\downarrow n_\uparrow}{\sqrt{2}}\right) \ , \\
    \psi_{4}^{(1)} &= \sqrt{\frac23}\ \Sigma^0_\uparrow n_\uparrow +\sqrt{\frac13}\ \Sigma^-_\uparrow p_\uparrow \ , \\
    \psi_{5}^{(1)} &= -\sqrt{\frac{1}{6}}\  \Sigma^0_\uparrow n_\uparrow+\sqrt{\frac{1}{3}}\ \Sigma^-_\uparrow p_\uparrow  + \sqrt{\frac{1}{2}}\ \Lambda_\uparrow n_\uparrow \ , \\
    \psi_{6}^{(1)} &= -\sqrt{\frac{1}{6}}\ \Sigma^0_\uparrow n_\uparrow+ \sqrt{\frac{1}{3}}\ \Sigma^-_\uparrow p_\uparrow - \sqrt{\frac{1}{2}}\ \Lambda_\uparrow n_\uparrow \ .
    \end{align}
    By replacing operators $F_{1\uparrow} F_{2\uparrow}$ in $\psi_i^{(1)}$ with $(F_{1\uparrow} F_{2\downarrow} + F_{1\downarrow} F_{2\uparrow})/\sqrt{2}$ or $F_{1\downarrow} F_{2\downarrow}$, one gets $\psi_i^{(2)}$ or $\psi_i^{(3)}$.
    Scattering channels can be read off and are listed in Eqs.~(\ref{basis1})-(\ref{basis3}). To further identify the irrep of each $\psi_i$, we compute the eigenvalues of Casimir operators acting on these states. $SU(3)$ has two Casimir operators, 
    \bee
    \mathcal{C}_1 = \sum_a T^a T^a \ , \quad \mathcal{C}_2 = \sum_{abc} d^{abc} T^a T^b T^c \ ,
    \ee
	where $d^{abc}=2 {\rm Tr}(\{T^a, T^b\}T^c)$ is the totally symmetric $d$ constants of $SU(3)$. Here $\mathcal{C}_1 = (p^2 +q^2+3p+3q +pq)/3$ and $\mathcal{C}_2=(p-q)(3+p+2q)(3+q+2p)/18$, where $(p,q)$ for the irreps are: ${\bf 27}=(2,2), {\bf 10} =(3,0), {\bf \overline{10}}=(0,3), \bf{8}=(1,1)$, and $\bf{1}=(0,0)$ \cite{Georgi:2000vve}. Every irrep can be determined by a unique set of eigenvalues of $\mathcal{C}_1$ and $\mathcal{C}_2$. This process also allows us to find the values of $C_{\bm{27}}, C_{\bm{10}}, C_{\bm{\overline{10}}},  C_{\mathbf{8_S}}$ and $C_{\mathbf{8_A}}$ in terms of the $c_i$. We complete the tables below by repeating this process for every $(Q, S)$ sector. 
	Alternatively, the states can be found using lowering operators in $SU(3)$. Since these states are weights of each irrep, one can start by identifying the unique highest weight in each irrep, whose $(Q,S)$ values are fixed by the representation theory. To find the highest weights of $\bm{8}_S, \bm{8}_A$ and $\bm{1}$, it is useful to also consider the action of two Casimir operators, which fixes the linear combination of multiple physical states with same $(Q,S)$ values. Then one can derive all remaining weights by repeated action of three lowering operators, $I_-=T^1-iT^2, V_-=T^4-iT^5$, and $U_- = T^6-iT^7$.

	\begin{table}[h]
	\begin{tabular}{|>{\centering\arraybackslash}m{10cm}|}
		\hline 
		scattering channels    \\
		\hline
		$n n$ \\
		\hline
		$n p$ \\
		\hline
		$p p$ \\
		\hline
		$n \Sigma^- $ \\
		\hline
		$\sqrt{\frac 2 3}\Sigma^0 n+\sqrt{\frac 1 3} \Sigma^- p$   \\
		\hline
		$-\sqrt{\frac{1}{30}}\Sigma^0 n+\sqrt{\frac{1}{15}} \Sigma^- p + \sqrt{\frac{9}{10}} \Lambda n$   \\
		\hline
		$-\sqrt{\frac 1 3}\Sigma^+ n+\sqrt{\frac 2 3} \Sigma^0 p$    \\
		\hline
		$\sqrt{\frac{1}{15}}\Sigma^+ n+\sqrt{\frac{1}{30}} \Sigma^0 p + \sqrt{\frac{9}{10}} \Lambda p$   \\
		\hline
		$\Sigma^+ p$ \\
		\hline
		$\sqrt{\frac{1}{3}} \Sigma^+ \Sigma^-  - \sqrt{\frac{2}{3}}  \Sigma^0 \Sigma^0 $\\
		\hline 
		$\sqrt{\frac{3}{5}} \Lambda \Sigma^0 +\sqrt{\frac{1}{5}} \Xi^- p - \sqrt{\frac{1}{5}} \Xi^0 n$  \\
		\hline
		$\sqrt{\frac{1}{60}} \Sigma^+ \Sigma^-  - \sqrt{\frac{1}{120}}  \Sigma^0 \Sigma^0 - \sqrt{\frac{3}{20}} \Xi^0 n - \sqrt{\frac{3}{20}} \Xi^- p + \sqrt{\frac{27}{40}} \Lambda \Lambda$\\
		\hline
		$\Sigma^- \Sigma^-$ \\
		\hline
		$\Sigma^- \Sigma^0$ \\
		\hline            
		$\sqrt{\frac{3}{5}} \Lambda \Sigma^- - \sqrt{\frac{2}{5}} \Xi^- n$ \\
		\hline
		$\sqrt{\frac{3}{5}} \Lambda \Sigma^+ - \sqrt{\frac{2}{5}} \Xi^0 p$ \\
		\hline
		$\Sigma^+ \Sigma^0$ \\
		\hline
		$\Sigma^+ \Sigma^+$ \\
		\hline
		$\Sigma^- \Xi^- $ \\
		\hline
		$-\sqrt{\frac{2}{3}}\Sigma^0 \Xi^- +\sqrt{\frac 1 3} \Sigma^- \Xi^0$    \\
		\hline
		$-\sqrt{\frac{1}{30}}\Sigma^0 \Xi^- +\sqrt{\frac{1}{15}} \Sigma^- \Xi^0 + \sqrt{\frac{9}{10}} \Lambda \Xi^-$  \\
		\hline
		$\sqrt{\frac 1 3}\Sigma^+ \Xi^- +\sqrt{\frac 2 3} \Sigma^0 \Xi^0$    \\
		\hline
		$-\sqrt{\frac{1}{15}}\Sigma^+ \Xi^- +\sqrt{\frac{1}{30}} \Sigma^0 \Xi^0 + \sqrt{\frac{9}{10}} \Lambda \Xi^0$   \\
		\hline
		$\Sigma^+\Xi^0 $ \\
		\hline
		$\Xi^- \Xi^-$ \\
		\hline
		$\Xi^- \Xi^0$ \\
		\hline
		$\Xi^0 \Xi^0$\\
		\hline 
	\end{tabular}
	\caption{Baryon-baryon scattering channels in $\bm{27}$.}\label{channels-27}
\end{table}

\begin{table}[ht]
	\begin{tabular}{|>{\centering\arraybackslash}m{9cm}|}
		\hline 
		scattering channels \\
		\hline
		$\sqrt{\frac{3}{10}}\Sigma^0 n - \sqrt{\frac{3}{5}} \Sigma^- p + \sqrt{\frac{1}{10}} \Lambda n$   \\
		\hline
		$\sqrt{\frac{3}{5}}\Sigma^+ n+\sqrt{\frac{3}{10}} \Sigma^0 p - \sqrt{\frac{1}{10}} \Lambda p$   \\
		\hline
		$ - \sqrt{\frac{2}{5}} \Lambda \Sigma^0 +\sqrt{\frac{3}{10}} \Xi^- p - \sqrt{\frac{3}{10}} \Xi^0 n$  \\
		\hline
		$\sqrt{\frac{2}{5}} \Sigma^+ \Sigma^-  - \sqrt{\frac{1}{5}}  \Sigma^0 \Sigma^0 + \sqrt{\frac{1}{10}} \Xi^0 n + \sqrt{\frac{1}{10}} \Xi^- p + \sqrt{\frac{1}{5}} \Lambda \Lambda$\\
		\hline
		$\sqrt{\frac{2}{5}} \Lambda \Sigma^- + \sqrt{\frac{3}{5}} \Xi^- n$ \\
		\hline
		$\sqrt{\frac{2}{5}} \Lambda \Sigma^+ + \sqrt{\frac{3}{5}} \Xi^0 p$ \\
		\hline
		$\sqrt{\frac{3}{10}}\Sigma^0 \Xi^- +\sqrt{\frac{3}{5}} \Sigma^- \Xi^0 - \sqrt{\frac{1}{10}} \Lambda \Xi^-$  \\
		\hline
		$\sqrt{\frac{3}{5}}\Sigma^+ \Xi^- - \sqrt{\frac{3}{10}} \Sigma^0 \Xi^0 + \sqrt{\frac{1}{10}} \Lambda \Xi^0$   \\
		\hline
	\end{tabular}
	\caption{Baryon-baryon scattering channels in $\bm{8}_S$.}\label{channels-8S}
\end{table}

\begin{table}[]
	\begin{tabular}{|>{\centering\arraybackslash}m{8cm}|}
		\hline 
		scattering channels    \\
		\hline
		$\frac{1}{2} \Sigma^+ \Sigma^-  + \sqrt{\frac{1}{8}}  \Sigma^0 \Sigma^0 + \frac{1}{2} \Xi^0 n + \frac{1}{2} \Xi^- p + \sqrt{\frac{1}{8}} \Lambda \Lambda$\\
		\hline
	\end{tabular}
	\caption{Baryon-baryon scattering channels in $\bm{1}$.}\label{channels-1}
\end{table}

\begin{table}[h]
	\begin{tabular}{|>{\centering\arraybackslash}m{7cm}|}
		\hline 
		scattering channels    \\
		\hline
		$n p$ \\
		\hline
		$-\sqrt{\frac{1}{6}}\Sigma^0 n+\sqrt{\frac{1}{2}} \Sigma^- p + \sqrt{\frac{1}{2}} \Lambda n$   \\
		\hline
		$\sqrt{\frac{1}{3}}\Sigma^+ n+\sqrt{\frac{1}{6}} \Sigma^0 p - \sqrt{\frac{1}{2}} \Lambda p$   \\
		\hline
		$\sqrt{\frac{1}{2}} \Lambda \Sigma^- - \sqrt{\frac{1}{6}} \Sigma^- \Sigma^0 - \sqrt{\frac{1}{3}} \Xi^- n$ \\
		
		\hline
		$\sqrt{\frac{1}{2}} \Lambda \Sigma^+ - \sqrt{\frac{1}{6}} \Sigma^+ \Sigma^0 - \sqrt{\frac{1}{3}} \Xi^0 p$ \\
		\hline
		$\Sigma^- \Xi^- $ \\
		\hline
		$-\sqrt{\frac{2}{3}}\Sigma^0 \Xi^- +\sqrt{\frac 1 3} \Sigma^- \Xi^0$    \\
		\hline
		$\sqrt{\frac 1 3}\Sigma^+ \Xi^- +\sqrt{\frac 2 3} \Sigma^0 \Xi^0$    \\
		\hline
		$\Sigma^+\Xi^0 $ \\
		\hline
	\end{tabular}
	\caption{Baryon-baryon scattering channels in $\overline{\bm{10}}$.}\label{channels-10}
\end{table}

\begin{table}[h]
	\begin{tabular}{|>{\centering\arraybackslash}m{7cm}|}
		\hline 
		scattering channels    \\
		\hline
		$n \Sigma^- $ \\
		\hline
		$\sqrt{\frac 2 3}\Sigma^0 n+\sqrt{\frac 1 3} \Sigma^- p$   \\
		\hline
		$-\sqrt{\frac 1 3}\Sigma^+ n+\sqrt{\frac 2 3} \Sigma^0 p$    \\
		\hline
		$\Sigma^+ p$ \\
		\hline
		$\sqrt{\frac{1}{2}} \Lambda \Sigma^- + \sqrt{\frac{1}{6}} \Sigma^- \Sigma^0 + \sqrt{\frac{1}{3}} \Xi^- n$ \\
		\hline
		$\sqrt{\frac{1}{2}} \Lambda \Sigma^+ + \sqrt{\frac{1}{6}} \Sigma^+ \Sigma^0 + \sqrt{\frac{1}{3}} \Xi^0 p$ \\
		\hline
		$-\sqrt{\frac{1}{6}}\Sigma^0 \Xi^- +\sqrt{\frac{1}{3}} \Sigma^- \Xi^0 - \sqrt{\frac{1}{2}} \Lambda \Xi^-$  \\
		\hline
		$-\sqrt{\frac{1}{3}}\Sigma^+ \Xi^- +\sqrt{\frac{1}{6}} \Sigma^0 \Xi^0 + \sqrt{\frac{1}{2}} \Lambda \Xi^0$   \\
		\hline
		$\Xi^- \Xi^0$ \\
		\hline 
	\end{tabular}
	\caption{Baryon-baryon scattering channels in $\bm{10}$.}\label{channels-10bar}
\end{table}

\begin{table}[h]
	\begin{tabular}{|>{\centering\arraybackslash}m{7cm}|}
		\hline 
		scattering channels \\
		\hline
		$-\sqrt{\frac{1}{6}}\Sigma^0 n+\sqrt{\frac{1}{2}} \Sigma^- p + \sqrt{\frac{1}{2}} \Lambda n$   \\
		\hline
		$\sqrt{\frac{1}{3}}\Sigma^+ n+\sqrt{\frac{1}{6}} \Sigma^0 p - \sqrt{\frac{1}{2}} \Lambda p$   \\
		\hline
		$- \sqrt{\frac{2}{3}} \Sigma^- \Sigma^0 + \sqrt{\frac{1}{3}} \Xi^- n$ \\
		\hline
		$\sqrt{\frac{1}{2}} \Xi^0 n + \sqrt{\frac{1}{2}} \Xi^- p$\\
		\hline
		$ - \sqrt{\frac{1}{6}} \Xi^0 n + \sqrt{\frac{1}{6}} \Xi^- p + \sqrt{\frac{2}{3}} \Sigma^- \Sigma^+ $\\
		\hline
		$- \sqrt{\frac{2}{3}} \Sigma^+ \Sigma^0 + \sqrt{\frac{1}{3}} \Xi^0 p$ \\
		\hline
		$-\sqrt{\frac{1}{6}}\Sigma^0 \Xi^- +\sqrt{\frac{1}{3}} \Sigma^- \Xi^0 + \sqrt{\frac{1}{2}} \Lambda \Xi^-$  \\
		\hline
		$-\sqrt{\frac{1}{3}}\Sigma^+ \Xi^- +\sqrt{\frac{1}{6}} \Sigma^0 \Xi^0 - \sqrt{\frac{1}{2}} \Lambda \Xi^0$   \\
		\hline
	\end{tabular}
	\caption{Baryon-baryon scattering channels in $\bm{8}_A$.}\label{channels-8A}
\end{table}
	\clearpage
	
	\bibliographystyle{apsrev4-1}
	\bibliography{DOE_NP_ref}

\end{document}